\documentclass[%
 reprint,
 amsmath,amssymb,
 aps,
]{revtex4-2}

\usepackage{graphicx}
\usepackage{dcolumn}
\usepackage{bm}

\usepackage{chemist}
\usepackage{tikz}
\usepackage{array}
\usepackage{amsmath,amsfonts,amssymb}
\usepackage{eurosym}
\usepackage{xcolor}
\usepackage{braket}
\usepackage{float}
\usepackage{svg}
\usepackage{soul}
\usepackage{hyperref}
\usepackage{caption}
\usepackage{subcaption}
\usepackage[app]{overpic}
\usepackage{physics}
\usetikzlibrary{quantikz}

\begin{document}

\preprint{APS/123-QED}

\title{CALCULATING THE GROUND STATE ENERGY OF BENZENE UNDER SPATIAL DEFORMATIONS WITH NOISY QUANTUM COMPUTING}

\newcommand{\wsc}[1]{\textcolor{blue}{#1}} 
\newcommand{\wsi}[1]{\textcolor{blue}{\textit{#1}}} 
\newcommand{\wss}[1]{\textcolor{blue}{\st{#1}}} 

\newcommand{\mr}[1]{\textcolor{red}{#1}} 
\newcommand{\mri}[1]{\textcolor{red}{\textit{#1}}} 
\newcommand{\mrs}[1]{\textcolor{red}{\st{#1}}} 
\author{Wassil Sennane$^1$}
\email{wassil.sennane@totalenergies.com}
\author{Jean-Philip Piquemal$^2$}
\email{jean-philip.piquemal@sorbonne-universite.fr}
\author{Marko J. Ran\v{c}i\'{c}$^1$}
 \email{marko.rancic@totalenergies.com}
\affiliation{1. TotalEnergies, 8, Boulevard Thomas Gobert, Nano-INNOV – Bât. 861,– 91120 Palaiseau – France \\ 2. Laboratoire de Chimie Théorique, Sorbonne Université, UMR7616 CNRS, Paris - France.}

\begin{abstract}
In this manuscript, we calculate the ground state energy of benzene under spatial deformations by using the variational quantum eigensolver (VQE). The primary goal of the study is estimating the feasibility of using quantum computing ansatze on near-term devices for solving problems with large number of orbitals in regions where classical methods are known to fail. Furthermore, by combining our advanced simulation platform with real quantum computers, we provided an analysis of how the noise, inherent to quantum computers, affects the results. The centers of our study are the hardware efficient and quantum unitary coupled cluster ansatze (qUCC). First, we find that the hardware efficient ansatz has the potential to outperform mean-field methods for extreme deformations of benzene. However, key problems remain at equilibrium, preventing real chemical application. Moreover, the hardware efficient ansatz yields results that strongly depend on the initial guess of parameters - both in the noisy and noiseless cases - and optimization issues have a higher impact on their convergence than noise. This is confirmed by comparison with real quantum computing experiments. On the other hand, the qUCC ansatz alternative exhibits deeper circuits. Therefore, noise effects increase and are so extreme that the method never outperform mean-field theories. Our dual simulator/8-16 qubits QPU computations of qUCC appears to be a lot more sensitive to hardware noise than shot noise, which give further indications about where the noise-reduction efforts should be directed towards. Finally, the study shows that qUCC method better captures the physics of the system as the qUCC method can be utilized together with the Hückel approximation. We discussed how going beyond this approximation sharply increases the optimization complexity of such a difficult problem.

\end{abstract}

\maketitle


\section{Introduction}

Quantum computing opens a new era of calculations thanks to quantum superposition and quantum entanglement. While classical computers handle binary information, quantum computers use entangled superposition of states as information carriers. Some algorithms, such as the well-known Grover algorithm \cite{grover1996fast} and Shor's algorithm \cite{shor1999polynomial}, will bring new efficient ways of solving complex problems if implemented on quantum computers. Moreover, the deepest understanding of complex molecules will become possible \cite{abrams1999quantum} with new techniques such as the Quantum Phase Estimation (QPE) algorithm \cite{mohammadbagherpoor2019improved} \cite{aspuru2005simulated} and the Variational Quantum Eigensolver algorithm (VQE) \cite{fedorov2022vqe} \cite{peruzzo2014variational} \cite{parrish2019quantum} \cite{bharti2021iterative} \cite{liu2019variational} \cite{nakanishi2019subspace} \cite{fujii2020deep} \cite{garcia2018addressing} \cite{cerezo2020variational} \cite{wang2019accelerated}. This new technology uses very fragile entangled states of the matter, which makes the management of multiple qubits without error correction difficult. As QPE requires quantum error correction we focus solely on the VQE in this paper.

There are in principle two approaches in which quantum computing is foreseen to bring value: through development of quantum inspired algorithms which are executed on quantum simulators and through algorithms which are executed on actual quantum hardware. Quantum simulators exploit advanced supercomputing platforms to emulate quantum-computing like environments. On the other hand the bottleneck in executing algorithms on actual hardware is quantum noise processes: qubit dephasing, qubit relaxation and readout errors. With two qubit error rates on the order of $1\%$ the maximal number of two qubit gates applied in a circuit is a few hundred \cite{temme2017error} \cite{li2017efficient}. 

An open question remains: could any of these two approaches bring immediate value in treating problems beyond for instance very simple molecules? We try to address these questions by solving the problem of benzene, a cyclic molecule with the formula \chemform{C_6H_6}.

\begin{figure*}[t!]
    \begin{minipage}{0.45\textwidth}
        \centering
        \includegraphics[scale=0.5]{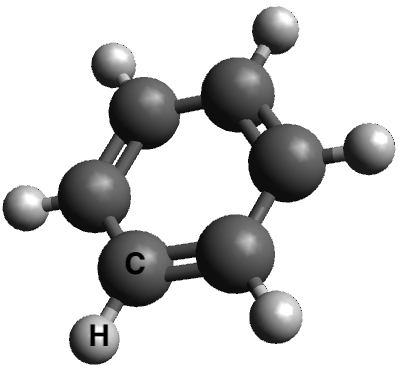}
        \put(-200,150){(a)}
        \put(20,150){(b)}
    \end{minipage}
    \begin{minipage}{0.5\textwidth}
        \centering
        \includegraphics[scale=0.5]{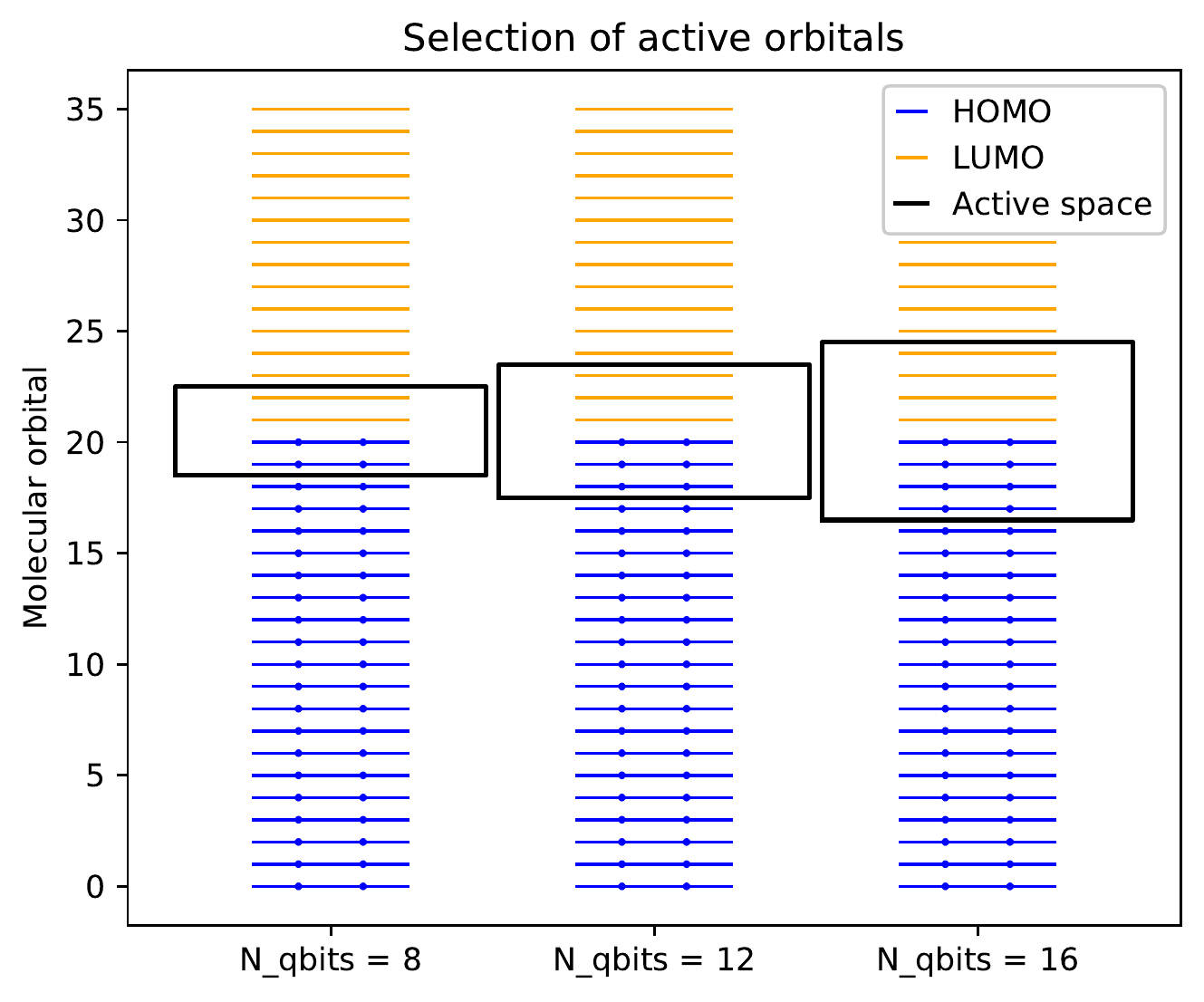}
    \end{minipage}
    \caption{(a) Molecular geometry of Benzene. (b) Selections of the active orbital space used throughout the manuscript.}
    \label{choice_orb}
\end{figure*}

In order to minimize the number of qubits required for a computation, we chose to work with the least computationally demanding basis - the sto-3g. In addition to the computational cost savings, this minimal basis set will amplify the errors, which can helps us identify the shortcomings of the methods we use. In this basis, each \chemform{H} is represented with a $1s$ orbital, and each \chemform{C} is represented with $1s, 2s, 2p_x, 2p_y, 2p_z$ \cite{lewars2011computational}. Therefore, the treatment of the entire molecule would require taking into account 36 orbitals, equivalently 72 spin orbitals for which 72 qubits would be required. Because simulating 72 qubits is beyond the simulation capabilities of our in-house simulator, capable of simulating up to 35 qubits, our approach relies on a reduction of the size of the system via active space selection. 132 qubits would have been required with 6-31g basis, and 228 qubits with cc-pVDZ basis set.

In this paper, we considered four ways of selecting the active orbital case. Figure \ref{choice_orb} (b) shows the active spaces used in this manuscript, all of which have an equal number of Highest Occupied Molecular Orbitals (HOMO) and Lowest Unoccupied Molecular Orbitals (LUMO). Intensive numerical testing showed that those active spaces with an equal number of HOMO and LUMO orbitals lead to lower energies as opposed to active spaces with unequal numbers of HOMO and LUMO orbitals. 

In order to test the added value of contemporary quantum computing methodologies we focus on situations where mean-field theories such as Hartree-Fock and density-functional theory are not precise enough \cite{google2020hartree} \cite{sokolov2020quantum}. Furthermore, we search for situations where classical coupled cluster ansatze are expected to yield non-physical solutions and full-configuration interaction calculations are, although theoretically possible, quite impractical due to a large active space \cite{sokolov2020quantum} \cite{large_FCI_1} \cite{rossi1999full}. Such situations occur when systems with a large number of orbitals are subject to spatial deformations further refereed to as distortions. We chose three types of such distortions. For each of these distortions, we will compute the Hamiltonian of the molecule in the 3 cases described in Figure \ref{choice_orb} (b) and calculate their lowest eigenvalue, which corresponds to the ground state energy of the size-reduced system. 

\begin{figure*}[t!]
    
    \includegraphics[scale=0.55]{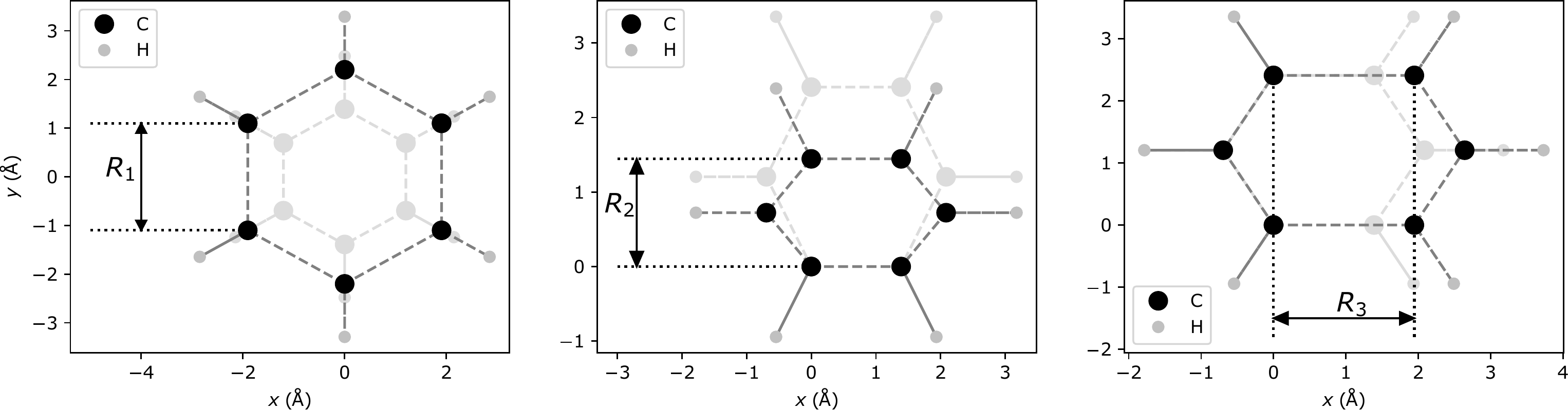}
    \caption{The 3 types of distortions studied. In the remainder of the manuscript these distortions are refereed to as distortion 1, distortion 2 and distortion 3 respectively.}
    \label{desc_dist}
\end{figure*}

Figure \ref{desc_dist} describes the three types of distortions we use in this paper. All the distortions studied here have the same spacing between one carbon and its closest neighboring hydrogen atom. This distance is fixed at $1.09$ {\AA} as in the equilibrium conformation of benzene. Generally, all fixed parameters have their equilibrium value. The first distortion is a uniform deformation of the molecule, in which the distance between two neighboring carbon atoms $R_1$ varies. In the second distortion, as Figure \ref{desc_dist} shows, the variable parameter is the distance $R_2$ between two opposite sides of the hexagon. The distance between the carbon atoms which belong each of these sides do not vary. Moreover, the two carbon atoms which do not belong to one of these sides have their spacing fixed. During the distortion, these two carbon atoms are vertically halfway of the two sides. For the third distortion, the idea is to divide the benzene into 2 identical triplets of carbon-hydrogen pairs. Then, these two parts are laterally moved one from another by varying the parameter $R_3$.

TotalEnergies is in possession of the Quantum Learning Machine (QLM) \cite{haidar2022open}, which allows large-scale simulations of quantum processing units (QPUs). This platform, operating 192 cores and 3 TB of RAM, is also compatible with QPUs from various providers such as IBM's quantum computers. Moreover, it should be noted that both myQLM is an open source version of the QLM libraries used and the codes for this study are public \cite{NEASQC}.

This manuscript is organized as follows : in Section \ref{sec:level01} we describe the methodolgy, then we show and discuss all our simulations in Section \ref{sec:level02}, moving with increased level of complexity going from noiseless simulations to compare different ansatze and identify optimizations issues, introducing shot noise and finally comparing simulated noise to real experiments in IBMQ. After all we conclude in Section \ref{sec_conc}.

\section{Methods\label{sec:level01}}

\subsection{\label{sec:level20}State-of-the-art classical computing methodologies}

A number of classical computational chemistry methods was performed in order to benchmark the quantum computing  results : Hartree–Fock (HF), second order Møller–Plesset perturbation theory (MP2), Coupled Cluster Single and Double(CCSD) and Triple(CCSDT), Configuration Interaction with Single and Double excitations (CISD), and Configuration Interaction using a Perturbative Selection (CIPSI). With these classical methodologies, the benzene system in sto-3g basis can be treated without orbital freezing. There are many possible open-source realizations for the listed methods. In our case for the HF, MP2, CCSD, CCSDT and CISD methods, the PySCF package version 1.7.5.1 was used \cite{sun2018pyscf,sun2020recent} on TotalEnergies in-house HPC architecture. Otherwise, the CIPSI method is the current state-of-the-art methodology for calculating energy levels of benzene \cite{loos2020performance}. It is implemented in Quantum package (QP) in the Master clusters at the Laboratoire de Chimie théorique (LCT). An open source implementation of wave function quantum chemistry methods was used \cite{garniron2019quantum}, mainly developed at the Laboratoire de Chimie et Physique quantiques (LCPQ) in Toulouse (France), and (LCT) in Paris. The CIPSI codes were executed in an efficient parallel fashion.

\subsection{\label{sec:level21}Orbital freezing}

Chemically, the lowest orbitals will be frozen and considered as full, they will be taken into consideration in the core energy, while the highest orbitals will be considered as non-active and empty. As Hückel's approximation suggests \cite{huckel1931quanstentheoretische}, most of the electronic behavior of aromatic molecules such as benzene is described with the $\pi$ orbitals of the carbon atoms, so we try to keep them active while reducing the active space. Since the two highest occupied orbitals are degenerate in benzene, the smallest case we consider is 4 electrons in 4 orbitals, which corresponds to the 8 qubits case in Fig \ref{choice_orb} (b). The 12 qubits case considers all $\pi$ orbitals active, while our 16 qubits case goes beyond the Hückel approximation. Mathematically, our in-house simulator the Quantum Learning Machine (QLM) \cite{haidar2022open} defines two subspaces of indices corresponding to active orbitals ($\mathcal{A}$) and occupied orbitals ($\mathcal{O}$). By computing the eigenvalues of the reduced density-matrix of the molecule, one obtains the Natural Orbital Occupation Numbers $n_i$ for each molecular orbital $i$ of the whole system. Then, the two subspaces are filled by a built-in function of the QLM which requires upper and lower thresholds $\epsilon_1$ and $\epsilon_2$ to select the orbitals :

\begin{equation}
    \mathcal{A} = \{i \ | \ n_i \in [\epsilon_2,2-\epsilon_1] \} \cup \{i \ | \ n_i \geq 2 - \epsilon_1, \  2(i+1) \geq N_{elec}  \} \nonumber
\end{equation}
\begin{equation}
    \mathcal{O} = \{i \ | \ n_i \geq 2 - \epsilon_1, \  2(i+1) < N_{elec}  \}.
\end{equation}

Once one has these subspaces, the built-in function performs the update of the one-body term and the core energy 

\begin{eqnarray}
\forall p,q \in \mathcal{A}, & & \nonumber \\ h_{pq} &\rightarrow& h_{pq} + \sum_{i \in \mathcal{O}} 2h_{ipqi} - h_{ipiq}, \\ E_{\rm core} &\rightarrow& E_{\rm core} + \sum_{i \in \mathcal{O}} h_{ii} + \sum_{i,j \in \mathcal{O}} 2h_{ijji} - h_{ijij}. \nonumber
\end{eqnarray}
Finally, the choice of the thresholds implies directly the choice of the active orbitals.

\subsection{\label{sec:level22}The Variational Quantum Eigensolver}

\begin{figure*}[t!]
     \centering
     \begin{subfigure}[b]{0.45\textwidth}
         \centering
         \includegraphics[width=\textwidth]{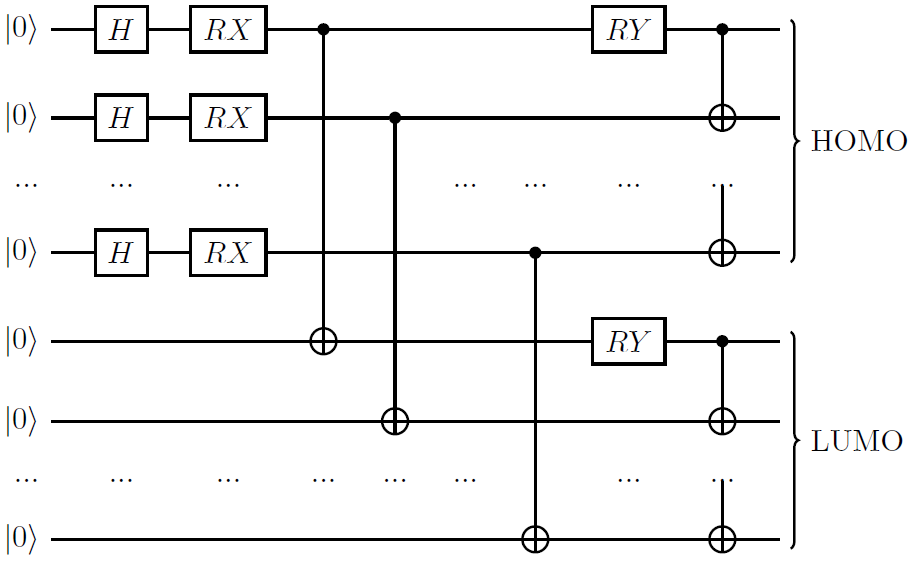}
         \caption{}
     \end{subfigure}
     \hfill
     \begin{subfigure}[b]{0.5\textwidth}
         \centering
         \includegraphics[width=\textwidth]{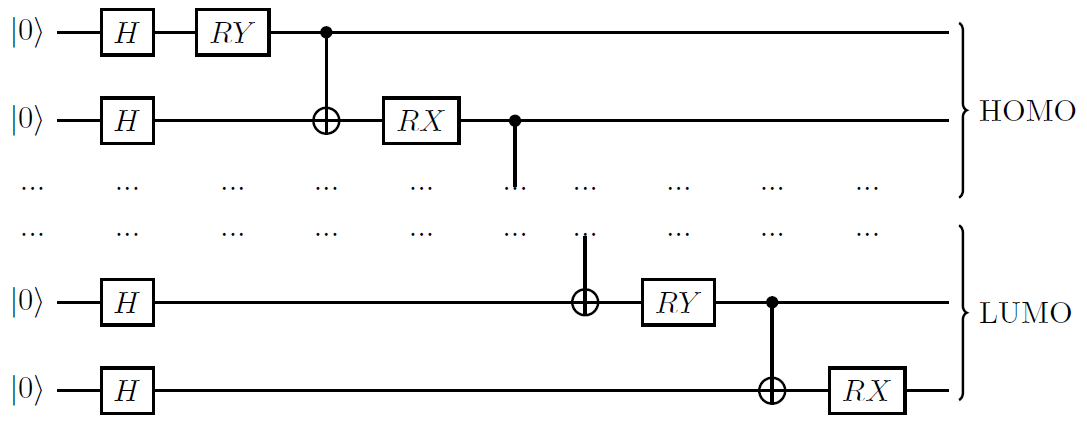}
         \caption{}
     \end{subfigure}
     \begin{subfigure}[b]{0.77\textwidth}
         \centering
         \includegraphics[width=\textwidth]{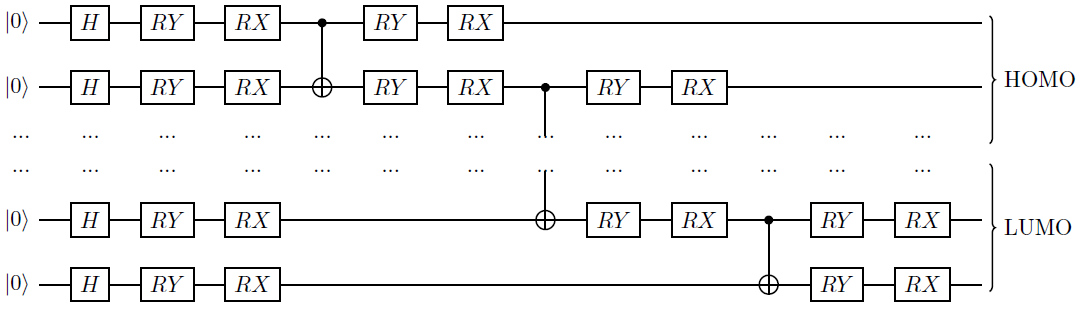}
         \caption{}
     \end{subfigure}
    \caption{Hardware efficient circuits used in this study (a) v1, (b) v2 and (c) v3. Each $RY$ and $RX$ gates are parameterized by an angle $\theta$. Here the three circuits have their depth $d=1$. The number of parameters in (a) and (b) is $dN_{qubits}$, while it is $2d(3N_{qubits}-2)$ for (c).} 
    \label{he_circ_345}
\end{figure*}

The Variational Quantum Eigensolver (VQE) \cite{fedorov2022vqe} \cite{peruzzo2014variational} \cite{parrish2019quantum} \cite{bharti2021iterative} \cite{liu2019variational} \cite{nakanishi2019subspace} \cite{fujii2020deep} \cite{garcia2018addressing} \cite{cerezo2020variational} \cite{wang2019accelerated} algorithm gives an estimation of the lowest eigenvalue of an eigenproblem, based on an optimization loop. First, one starts with a molecule, whose one-body and two-body terms are calculated by a chemistry package such as for instance PySCF \cite{sun2018pyscf}. Second, one applies a qubit-mapping transformation, such as the Jordan-Wigner transformation (see Appendix \ref{appendix_JW}) in order to obtain $H_{JW}$. This new Hamiltonian is written as a sum of Pauli strings, so it is understandable and measurable by quantum computers. Third, one creates a unitary state $\ket{\psi(\bm{\text{$\theta$}})}$ which depends on a set of parameters, and then one computes:

\begin{equation}
E(\bm{\text{$\theta$}}) = \frac{ \bra{\psi (\bm{\text{$\theta$}}) } H_{JW} \ket{\psi (\bm{\text{$\theta$}}) } }{ \braket{ \psi (\bm{\text{$\theta$}})} } = \sum_{j} h_j \bra{\psi(\bm{\text{$\theta$}})} P_{j} \ket{\psi(\bm{\text{$\theta$}})}
\end{equation} here $P_j$ are tensor products of Pauli matrices (labeled as Pauli strings in \ref{appendix_JW}), and $h_j$ are weights obtained from the Jordan-Wigner transformation.
The result is then communicated to a classical computer, which returns a new set of parameters, so an expectation value of energy of the new trial state $\ket{\psi(\bm{\text{$\theta'$}})}$ that will be calculated again with the quantum computer. This loop occurs until the optimizer finds the minimum of the energy. Thanks to the Ritz variational principle \cite{Ritz1909}, the variational solution obtained by the quantum computer can not be lower than the exact energy $E_0$ :

\begin{equation}
\forall \bm{\text{$\theta$}}, \ \ \ E_0 \leq E(\bm{\text{$\theta$}}) = \bra{\psi(\bm{\text{$\theta$}})}H_{JW}\ket{\psi(\bm{\text{$\theta$}})}.
\end{equation}

In the remainder of this paper, a COBYLA optimizer with maximally $1000$ iterations will be used \cite{lavrijsen2020classical}.

A critical point in solving the eigenvalue problem with the VQE is the selection of the trial state. There are mainly two different types of trial states, known as Hardware Efficient (HE) ansatz and the quantum computer Unitary Coupled Cluster (qUCC) ansatz. The qUCC ansatz is a chemically inspired trial state, with a large number of quantum gates and a precise estimation of the ground-state energy \cite{moll2018quantum} \cite{mizukami2020orbital} \cite{lee2018generalized} \cite{taube2006new}. The HE ansatz is a trial state designed to be easily implementable on quantum computers \cite{moll2018quantum} \cite{kandala2017hardware}  \cite{gard2020efficient} \cite{tang2021qubit} \cite{rattew2019domain} but with a less precise estimation of the ground state energy.

The Hadamard gate is an integral part of the HE ansatz, enabling the creation of superimposed states. Another ingredient in the HE ansatz is the CNOT gate - a controlled operation which flips the $\ket{0}$ and $\ket{1}$ states if the controlled qubit is $\ket{1}$. The RX and RY gates are the unitary rotation operators around the $x$ and $y$ axis in the Bloch sphere, parameterized by an angle. Defined by the exponentiation of the $X$ and $Y$ Pauli matrices, with these two operators, one is able to cover the whole Bloch sphere. \cite{nielsen2002quantum}

The circuit in Figure \ref{he_circ_345} (a) is chemically inspired. If we suppose that the $N_{qubits}/2$ first qubits represents the HOMO orbitals, then the Hadamard gate creates an electron on each of them. Starting from this hypothesis, combining these gates with the first set of CNOT entanglements makes a HOMO-LUMO mixing, and then the second set makes a spin mixing. The RX and RY gates parameterize the circuit. The circuit in Figure \ref{he_circ_345} (b) is a circuit which is easily implementable on a quantum computer with nearest-neighbor two-qubit interactions (linear chain of qubits). An electron created in all the orbitals, and we use alternatively a RY or a RX gate to parameterize the occupancy of each qubit (orbital). The circuit in Figure \ref{he_circ_345} (c) is also connectivity inspired with more single qubit gates as compared to its counterpart in Figure \ref{he_circ_345} (b). We first create an electron on each orbital with the Hadamard gates. Then, knowing that the Hadamard gate is given by a product of $RY$ and $RX$ gates $H = RY(-\frac{\pi}{2})RX(\pi)$, the set of RY-RX gates directly weights the occupancy of each orbital. The CNOT entanglements produce a linear circuit, and after each entanglement we add a new set of RY-RX gate in order to prevent the errors due to non-commuting gates. In other words, this theoretically could enable the circuit to choose the best order of gates at a cost of an optimization function with a larger number of variational parameters.

In our case, the depth of the circuit is the number of parameterized layers one adds to the first set of Hadamard gates. In Figure \ref{he_circ_345}, the depth is equal to 1, and the circuits are displayed in the case of a $4$ qubit Hamiltonian.

For the implementation of the qUCC ansatz, the circuit is provided by built-in functions of the QLM. As an input one and two-body integrals are required, with the total number of electrons, the subspace of spin orbitals $\mathcal{O}$, and the Natural Orbital Occupations Numbers and the energies of the active orbitals. By computing $n_{e_{act}} = n_{e} - 2|\mathcal{O}|$, one has the number of active electrons, that is to say the number of electrons that are distributed over the active orbitals. Then, $\mathcal{A}$ is divided into two sub-spaces : $\mathcal{I'}$ which contains the unoccupied active orbitals and $\mathcal{O'}$ which contains the occupied active orbitals, and the anti-Hermitian operator is created:

\begin{eqnarray}
     \forall p,q,r,s \in \mathcal{I'}^{2}\times\mathcal{O'}^{2}, & & \nonumber \\
     T_{qUCC}(\bm{\text{$\theta$}}) &=& \sum_{pr} \theta_{p}^{r}(a_{p}^\dagger a_r - a_r^\dagger a_p) \\
     & + & \sum_{p > q, r > s} \theta_{pq}^{rs}(a_p^\dagger a_q^\dagger a_r a_s - a_r^\dagger a_s^\dagger a_p a_q), \nonumber
\end{eqnarray}
where $\bm{\text{$\theta$}}$ is an array of parameters. After that, the built-in functions give the ansatz $\ket{\psi(\bm{\text{$\theta$}})} = e^{{T_{qUCC}(\bm{\text{$\theta$}})}}\ket{\Phi_0}$, with $\ket{\Phi_0}$ being the state obtained with the Hartree-Fock method \cite{haidar2022open}. The operator is Trotterized up to first order \cite{haidar2022open} \cite{hatano2005finding}, which gives Pauli strings that are decomposed in single and two-qubit gates. Although the initial state would be the Hartree-Fock guess if $\bm{\text{$\theta_0$}} = 0$, the QLM calculates an improved $\theta_0$ which returns the Møller-Plesset second-order (MP2) as initial guess:

\begin{equation}
\label{eq_qUCC_t0}
\begin{split}
    \forall p,r \in \mathcal{I'}\times\mathcal{O'}&, \ \ \ \theta_{p}^{r} = 0 \\
    \forall p,q,r,s \in \mathcal{I'}^{2}\times\mathcal{O'}^{2}&, \ \ \ \theta_{pq}^{rs} = \frac{h_{pqrs}-h_{pqsr}}{\varepsilon_r+\varepsilon_s-\varepsilon_p+\varepsilon_q}
\end{split}
\end{equation} with $\varepsilon_i$ is the energy of the orbital $i$. As MP2 is a post-Hartree-Fock method, which takes into account some of the electronic correlation, the initial guess of the qUCC ansatz is expected to be better.

\subsection{\label{sec:level23}Noise models}

In this paper, we computed both noiseless and noisy simulations. The noisy simulations take into consideration two types of noise - shot noise and idle hardware noise. Shot noise originates from the final measurement of the qubit register. As we use a simulator of quantum computers, entire circuits are applied to the qubit register and then the machine has an access to the exact frequencies. Including shot noise means use a number of random shots to make an estimation of these frequencies and use them as a result. If one wants to measure the exact probabilities of any quantum state, an infinite number of shots is required. In a realistic experiment it is impossible to perform an infinite number of evaluations. Shot noise is a difference in measured probabilities between a finite and infinite number of realizations in any experiments in quantum computing. Our simulator is able to calculate these two cases by fixing the number of shots $n_{shots}$, which has to be $0$ to obtain the exact probabilities.

The hardware noise describes the decoherence through the execution of the circuit \cite{brune1996observing} \cite{degen2017quantum} \cite{frey2020simultaneous}. Within our in-house simulator two ways of incorporating noise exist - an ''idle noise" model and a full noise model. In the first one, the qubit gates are ideal and the noise appears only when the qubit is inactive. In the latter noise model decoherence and relaxation occurs both when the qubit is actively operated on and when the qubit is inactive. Due to the long simulation times of the full noise model in the remainder of this paper we will tackle noise from an idle perspective. 

Indeed, unlike pure quantum states which are well described with a statevector, open quantum systems must employ a density matrix description to capture the full noise dynamics. Giving a description of noise in an open-quantum system where the state of environment at time $t_1$ is correlated with the state of environment at time $t_2>t_1$ usually requires a solution of a generalized master equation for the density matrix such as the Zwanzig-Nakajima equation \cite{nakajima1958quantum, zwanzig1960ensemble}. As this integro-differential equation is \textit{per se} unsolvable for a realistic system, further approximations needs to be made such as the commonly employed Markov approximation. The Markov approximation assumes a memory-less environment, one where the state of the environment at time $t_1$ is totally uncorrelated with the state of environment at time $t_2$. Contemporary quantum computers, such as those from IBMQ, have a so-called $1/f$ noise spectral density, where the noise is a mixture of predominantly non-Markovian and in a smaller fraction of Markovian noise \cite{krantz2019quantum, martina2022learning}. However, the Lindblad-like equation \cite{breuer2002theory} which is the result of such approximations is a matrix partial-differential equation and is still difficult to solve for large systems. Consequently, our in-house simulator offers simplified quantum noise models.

\begin{figure*}[t!]
    \centering
    \includegraphics[width=\textwidth]{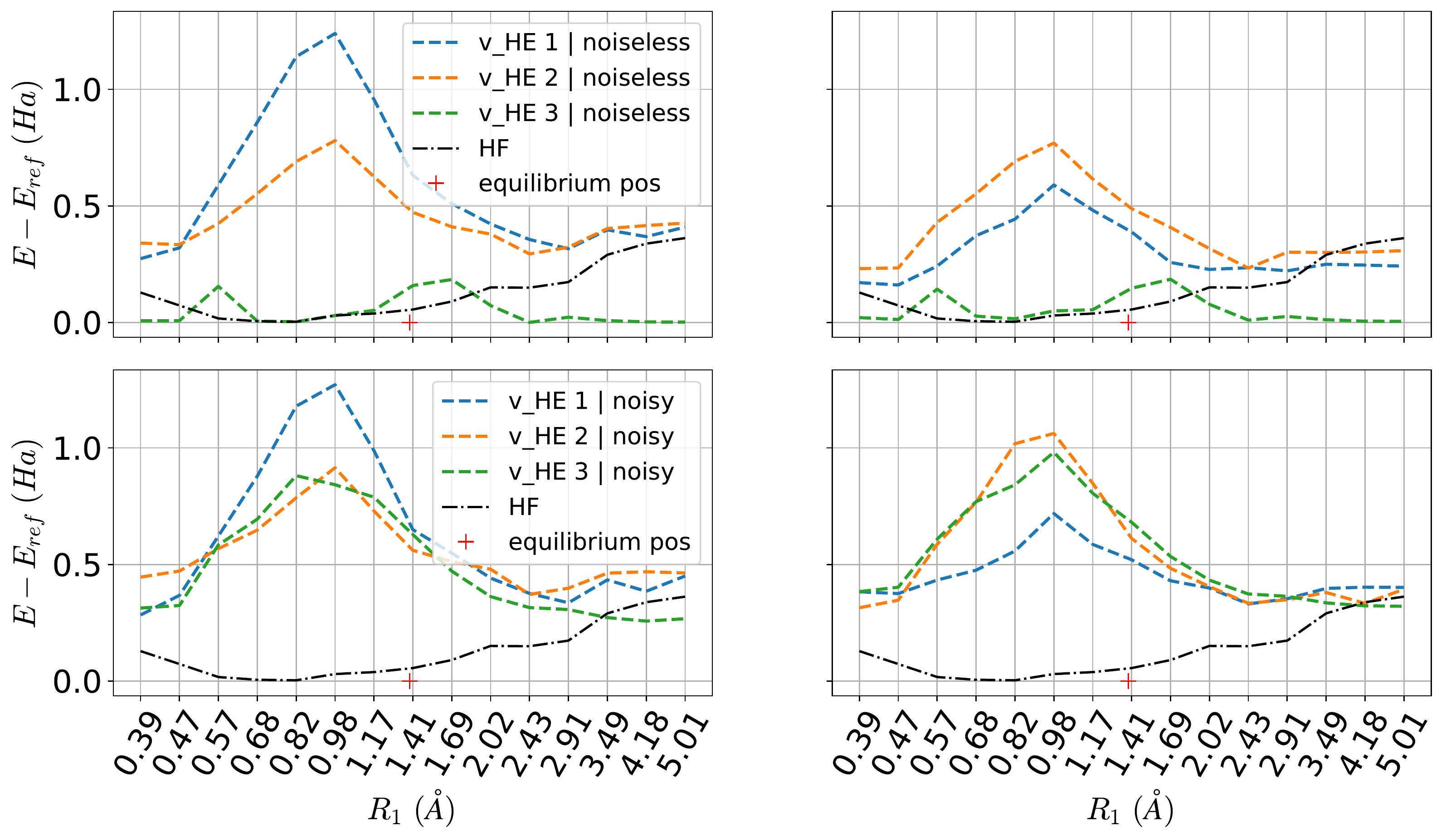}
    \caption{Difference between the ground state energy curves of $8$ qubits system obtained with HE ansatz and the reference obtained with full diagonalization of $8$ qubits system. All the figures have the same y-axis. left: $d=1$ right: $d=2$, top: noiseless, bottom: noisefull.}
    \label{compHEdist_full}
\end{figure*}

The QLM allows the incorporation of this model by applying amplitude damping and pure dephasing, with built-in functions which are returned by the noisy quantum processing unit simulator. By manipulating the density matrix describing the qubits with Kraus operators, the dephasing is implemented with a characteristic decay law on its anti-diagonal terms, while the amplitude damping affects both the diagonal and the anti-diagonal terms with a similar decay law (see Appendix \ref{appendix_idle_noise}). The built-in functions of the QLM require $T_1$ and $T_2$ - the two characteristic relaxation and dephasing times, alongside with gate duration. Table \ref{noisy_par} lists the values we used in the reminder of the paper.

\begin{table}[H]
\centering
\begin{tabular}{c|c|c|c|c|c|c|c|c}
\hline
$T_1$ & $T_2$ & $Z$ & $X$ & $Y$ & $RX$ & $RY$ & $RZ$ & CNOT\\
\hline
$50$ $\mu$s & $50$ $\mu$s & $60$ ns & $60$ ns & $60$ ns & $60$ ns & $60$ ns & $60$ ns & $150$ ns\\
\hline
\end{tabular}
\captionof{table}{Relaxation, dephasing and gate-duration times.}
\label{noisy_par}
\end{table}

\section{Results\label{sec:level02}}

\subsection{\label{sec:level24}Comparison of HE ansatz}

In this section we perform a comparison between the different HE circuits, for size-reduced cases with 4 electrons in 4 orbitals and for all three distortions. The first distortion is represented in Fig. \ref{compHEdist_full} (see Fig. \ref{compHEdist_full_appendix} in Appendix \ref{appendix_curves} for second and third distortions). First, for depth equal to $1$ and $2$, we display noiseless simulations, and then we display noisy simulations. These noisy simulations take into consideration shot noise and an idle noise model of the quantum processing unit. Moreover, while state-of-the-art post-Hartree-Fock solvers such as the Coupled-Cluster Single-Double excitation (CCSD) method is known to be very precise around the equilibrium position, they yield nonphysical solutions when the chemical system is distorted \cite{sokolov2020quantum} (and Fig. \ref{fig:compa}). Because of this and due to the fact that we want to make sure how much of two-body correlations does quantum computing capture we chose the Hartree-Fock energy to be the method which will be compared with our quantum simulations.

The energetic reference $E_{ref}$ is, for each dimension of each distortion, the exact ground state energy of the frozen-orbitals system, which is obtained with a full numerical diagonalization of the 8 qubit Hamiltonian. Since it is the target energy, we refer to this energy as "exact". Each point of the simulations represents an average of $50$ trials. In this work, a "trial" consists on choosing an initial guess and using it with VQE (with or without a model of noise).

\begin{figure*}[t!]
     \centering
     \includegraphics[width=\textwidth]{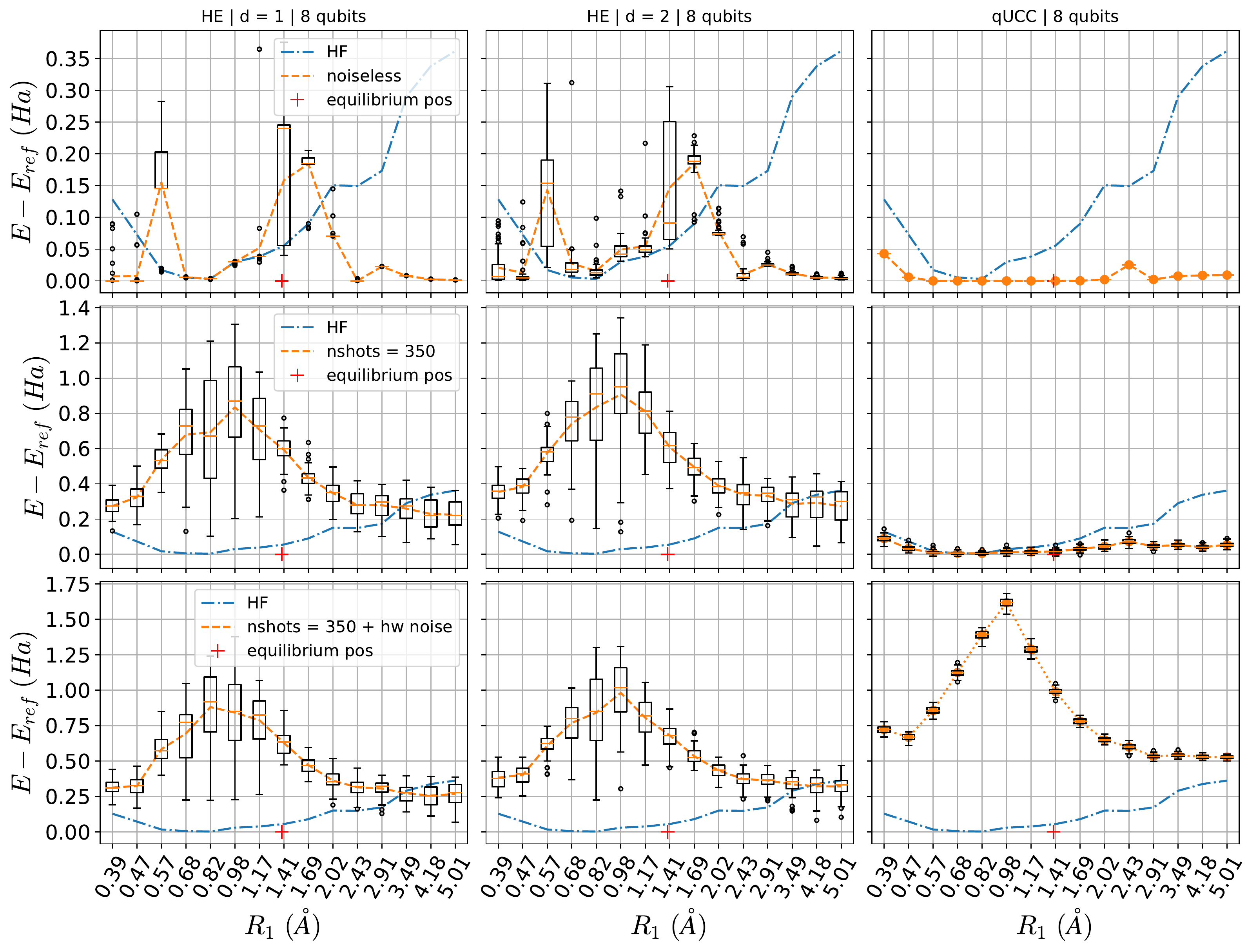}
     \caption{Difference between the ground state energies of 8 qubits system obtained with HE and qUCC ansatz and the reference obtained with full diagonalization of 8 qubits system. The boxplots give an overview of the amplitude of the variations due to optimization, shot noise and hardware noise issues. Each row share the same y-axis.}
     \label{9sub_dist1}
\end{figure*}

In the Figure \ref{compHEdist_full}, we can see that the v3 circuit yields a lower energy as compared to v1 and v2. The Hartree-Fock method on the other hand, works better around the equilibrium position, but when the distortion start to be extreme, the system is in a position far from equilibrium and this Hardware efficient circuit is has a supreme performance in calculating the ground state energy. The solution obtained with the v1 ansatz is improved with depth, while the ansatz v2 and v3 have their depth already optimal at $d=1$. As the number of parameters in v1 and v2 is $dN_{qubits}$, we can conclude that there is not obvious link between the number of parameters and the accuracy.

In the Figure \ref{compHEdist_full_appendix} (a), the v3 ansatz is still supreme to its HE counterparts but it can not manage to be closer than the Hartree-Fock energy. This difference with the distortion 1 comes from the distortion itself: as the Figure \ref{desc_dist} shows, in the case of distortion 1, when you change the distortion parameter, all the carbon atoms move. In the case of distortion 2, when you change the distortion parameter, there is two pairs of carbon atoms that have always the same spacing, so their correlation is the same than the equilibrium position correlation. This can also explain why the Hartree-Fock method provides a better approximation in the case of distortion 2 than distortion 1 - its deviation from the target value is always lower than $0.2$ $Ha$, while it reaches $0.4$ $Ha$ in distortion 1.

For the three distortions we can make the same general conclusions: the v3 circuit is supreme to v1 and v2, and the depth of the circuit impacts those results in a minor fashion. Around the equilibrium positions, the Hartree-Fock energy is more precise than these quantum methods. But, as we start to distort the system, the HE method gives better results. Eventually, these methods are rather complementary. Due to such a supreme performance of ansatz v3 we will use only this HE ansatz in the remaining subsections of the manuscript. The supreme performance of ansatz v3 is somewhat expected - it provides entanglement between neighboring qubits and has a large pool of single qubit gates.

\subsection{\label{sec:level25}HE vs qUCC}

\begin{figure*}[thb!]
     \centering
     \includegraphics[width=\textwidth]{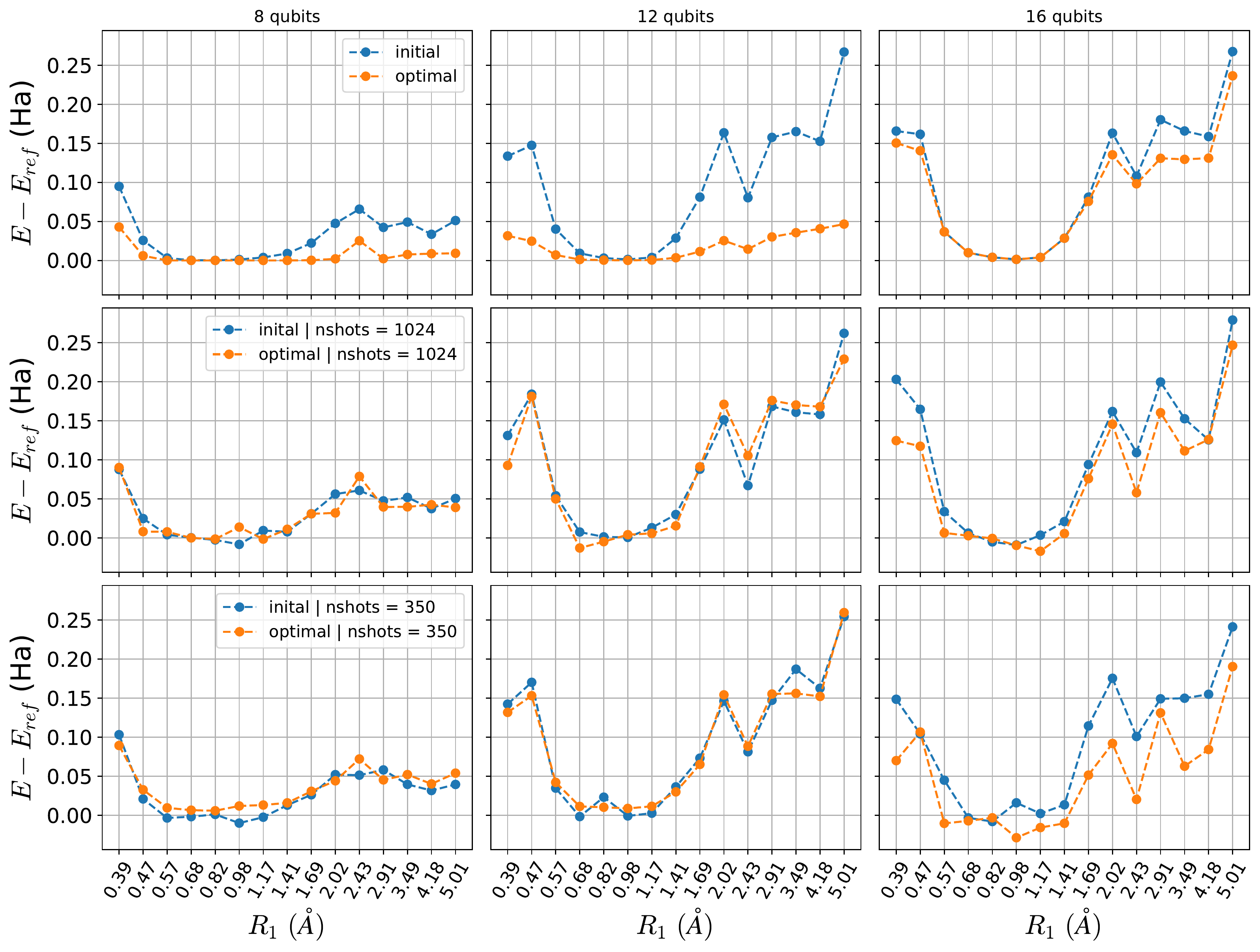}
     \caption{Comparison between initial and optimal guess yielded by qUCC ansatz, for the first distortion with $E_{ref}$ being obtained with full diagonalization of the active-space Hamiltonian, with corresponding number of qubits written on top of figures. All the figures share the same y-axis. The noise increases from top to bottom line.}
     \label{UCC_init_dist1}
\end{figure*}

In this subsection, we perform calculations in order to compare the HE method with different depths with qUCC method, in the case of 8 qubits systems for each distortion. The reference $E_{ref}$ is, for each length of each distortion case, the ground state energy of the 8 qubits system, obtained by a full diagonalization of the active-space Hamiltonians of 8 qubits. The optimizer is COBYLA with a maximum of $1000$ iterations.

For each of the Figures \ref{9sub_dist1}, \ref{9_sub_full_appendix} (a) and \ref{9_sub_full_appendix} (b) (see Appendix \ref{appendix_curves} for the second and third distortions), the two first columns represents the ground state energies obtained with the HE ansatz with a depth equals to $1$ and $2$. The third column shows the qUCC results. The first row displays noiseless simulations, while the two others show simulations with noise. The second row corresponds to the simulations with shot noise only and the third row refers to simulations with both shot noise and hardware noise. Finally, the HE results of the first and third row are the same than the results of Figure \ref{compHEdist_full}, then all the HE curves and the noisy qUCC curves displayed are the average of 50 trials, while the noiseless qUCC simulations are computed only one time.

The first thing that we draw the attention of the reader to is the presence of box-plots, which represents a variance in the multiple trials. The box denotes where $50 \%$ of the values converge to, while the full $100 \%$ is between the whiskers. The single points represents results that are beyond $1.5$ times the whisker size. This variance through the results is due to the choice of the initial guess, and the optimizer itself. For the initial guess of parameters in the HE method we choose a random vector. As every new trial requires a new initial guess, the starting trial state is never the same, so at the end, the algorithm can be confined in a local minima which depend heavily on the initial guess. This illustrates the shortcomings of the COBYLA optimizer, which is not robust enough to escape local minima and converge towards the global solution.

Generally, the noiseless simulations shows that the variance is higher around the equilibrium position. Moreover, the implementation of shot noise seems to equilibrate the variance across the results, and the consideration of hardware noise decrease the variance of the results, although this makes the HE method outperformed by classical methods.

Otherwise, the qUCC method has a very low variance, which can be explained by its accurate initial guess which seems to prevent the optimizer to converge towards a local minimum. In contrast to HE, the presence of hardware noise causes qUCC methods to give less good results compared to the HE method - a feature which is not too surprising given the large number of gates in the ansatz. It is also important to point out that qUCC is a lot more sensitive to the hardware noise than to the shot noise.

\subsection{\label{sec:level26}Optimization in qUCC}

In this part, we focus on the qUCC simulations. We compare the initial guess of qUCC with its value obtained after optimization in Fig. \ref{UCC_init_dist1} (see Fig. \ref{UCC_init_full} (a) and (b) in  Appendix \ref{appendix_curves} for the second and third distortions). The reference $E_{ref}$ is, for each length of each distortion case, the ground state energy of the $N_{qubits}$ qubits system (written on top of the columns), obtained with a full diagonalization of the active-space Hamiltonians of $N_{qubits}$ qubits. The optimizer is COBYLA with a maximum of $1000$ iterations. Each of the four columns of graphs corresponds to a different selection of the active orbital space in the system. Each row of graphs corresponds to a particular noise model. The first row displays noiseless simulations, while the two others show simulations with only shot noise, fixed at $1024$ for the second row and $350$ for the third row. The initial guesses are single simulations. For the 4, 8 and 12 qubit systems, the noisy optimal guess with $350$ shots displayed is an average of 50 trials. Except these, all the optimal guesses are obtained with only one trial. 

For each distortion, the 8 qubits case shows that the initial guess of qUCC is a really accurate guess. In particular for certain lengths in the first distortion, the difference between the initial guess, the optimal guess and the target energy is not perceptible at this scale. On the other hand, going from 4 to 12 qubits, the noiseless row illustrates that the difference between the initial guess and the optimal guess increases. However, when the number of qubits is 16 there is no substantial increase in precision between the initial guess and the optimal solution. This has to do with the fact that aromatic molecules such as benzene are well described with an active space of 6 orbitals (12 qubits). Indeed, the larger the system, the more parameters are added. The more parameters one adds, the more iterations one needs to converge. This lack of iterations explains also why the 16 qubits optimal guesses seems to be higher in energy than the 12 qubits ones. \label{12vs16}

Concerning the noisy simulations, the important information is that the optimal energy may be higher than the initial guess, which we found quite surprising. We understand that this phenomenon is caused by quantum noise. Indeed, let us imagine an optimization surface: increasing the level of noise decreases the smoothness of the optimization surface\cite{lavrijsen2020classical}, and the harder it is to reach the real minimum energy. Moreover, the optimization surface dynamically changes between realization - when the optimizer changes the parameters to check if it there is a better solution, the presence of noise can prevent the optimizer from recovering the previous ''better" solution. 

\subsection{\label{sec:level27}Increasing the number of qubits}

\begin{figure*}[thb!]
     \centering
     \begin{subfigure}[b]{\textwidth}
         \centering
         \includegraphics[width=\textwidth]{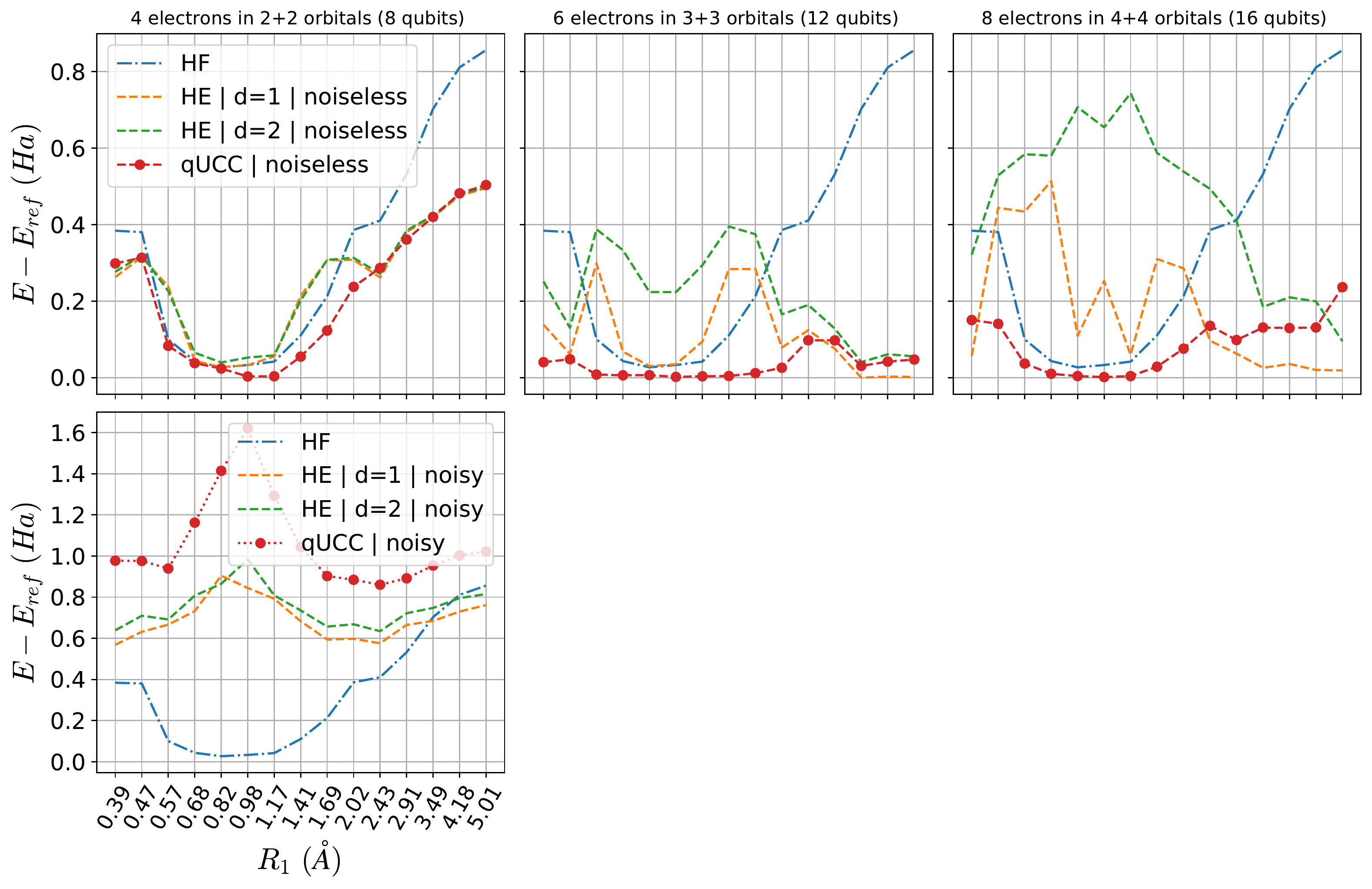}
         \label{full_dist1}
     \end{subfigure}
    \caption{Comparison of ground state energy curves of different number of qubits obtained with different methods, for the first distortion. $E_{ref}$ is the energy obtained with full diagonalization of the 16 qubits active-space Hamiltonian.  Each row share the same y-axis.}
    \label{full_dist_qb}
\end{figure*}

In this section we compare the results computed for the 4, 8, 12 and 16 qubits systems, for the three distortions in the Figures \ref{full_dist_qb}, \ref{full_dist_qb_B} (a) and \ref{full_dist_qb_B} (b). The first row displays the noiseless simulations, while the second row show the simulations with shot and hardware noise, up to 8 qubits due to capabilities of the QLM. Each column refers to a particular size of the system. The reference $E_{ref}$ is, for each length of each distortion case, the ground state energy of the $16$ qubits system, obtained with a full diagonalization of the active-space Hamiltonians of $16$ qubits. The optimizer is COBYLA with a maximum of $1000$ iterations. The qUCC noiseless simulations are obtained with one trial, while the noisy ones are an average of 50 trials. All the HE calculations are the results of 50 trials, except for 16 qubits systems where there is only one trial due to the large computational times required to obtain more than 1 trial.

Generally, for each distortion in the noiseless row, the qUCC method is very accurate in the case of noiseless simulations and it outperforms the classical methods. The Figures \ref{full_dist_qb}, \ref{full_dist_qb_B} (a) and \ref{full_dist_qb_B} (b) show that from 8 qubits systems to 12 qubits systems, the qUCC ansatz provides decreasing energies indicative of a better solution as a larger number of orbitals is activated.

The HE method has a variable precision depending on the distortion and the size of the system. As said before, the applicability of the HE ansatz is proportional to the level of distortion. The Figure \ref{full_dist_qb} shows explicitly that for extreme distortions, the HE method provides a better energy than the qUCC method, and does so with the same maximum number of iterations.

Likewise, when the deformation is less pronounced, the qUCC method and even the classical methods outperforms the HE method, which gives an increasing energy as the system grows. This phenomenon, which occurs also in the second and the third distortion, can also be due to the maximum number of iterations available. Indeed, for both HE and qUCC methods, the number of parameters increase with the number of qubits, therefore the convergence becomes more difficult with a growing system, while we have set our optimizer to maximally 1000 iterations throughout the manuscript. As in subsection \ref{sec:level26}, this explains why the qUCC method seems to stagnate between an active orbital selection of 12 and 16 qubits.

Concerning the noisy simulations, the three distortions have quite the same behavior. The qUCC method totally fails when one tries to include hardware noise, and even if the HE efficient method gives better results they are still sub-optimal because both are outperformed by classical methods. One can also notice that the quantum methods are further from the reference around the equilibrium position than for extreme distortions, in which quantum methods provide a better solution. \\

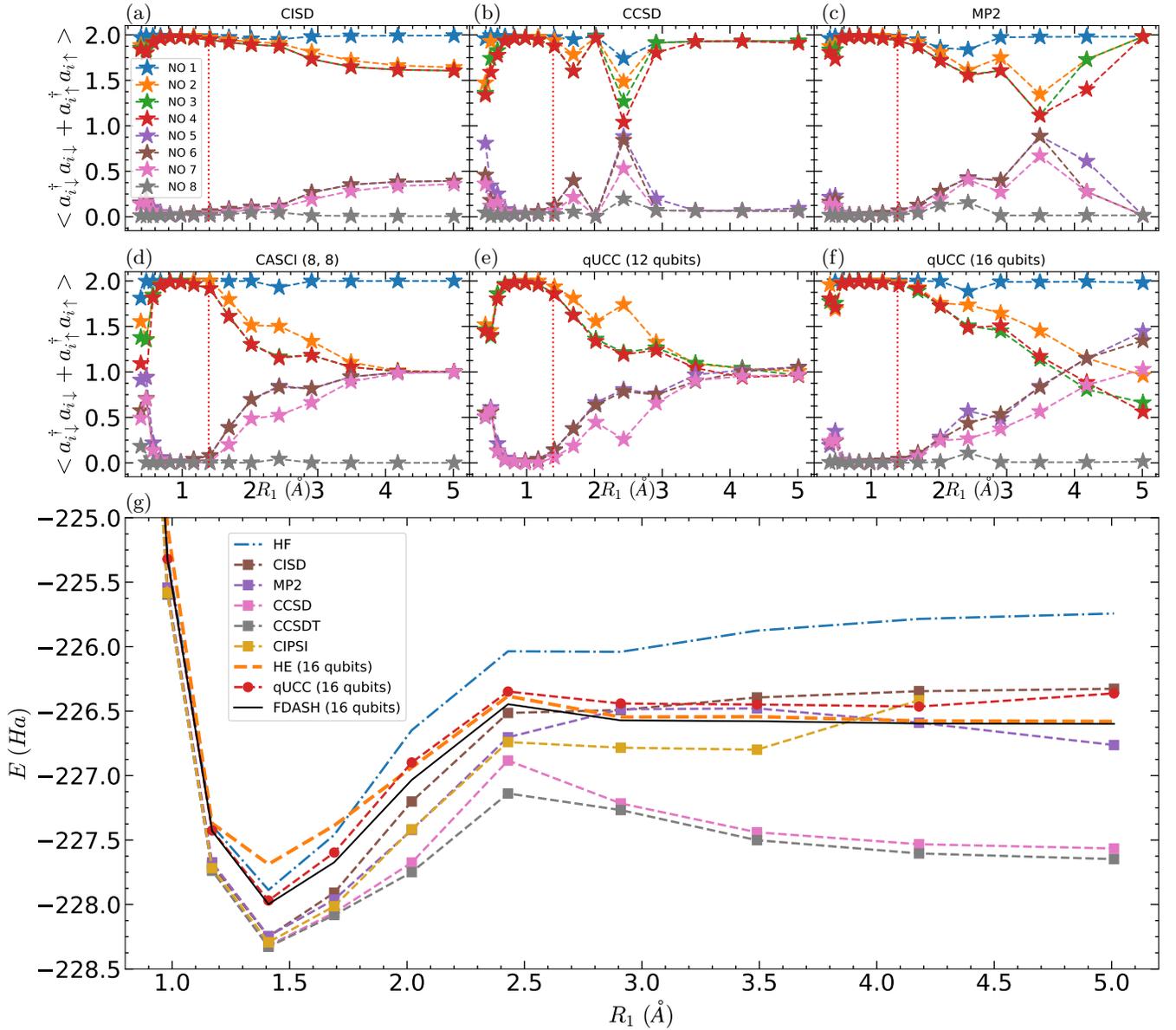
\begin{figure*}[t!]
     \begin{subfigure}[H]{\textwidth}
         \centering
         \begin{overpic}[width=1\textwidth]{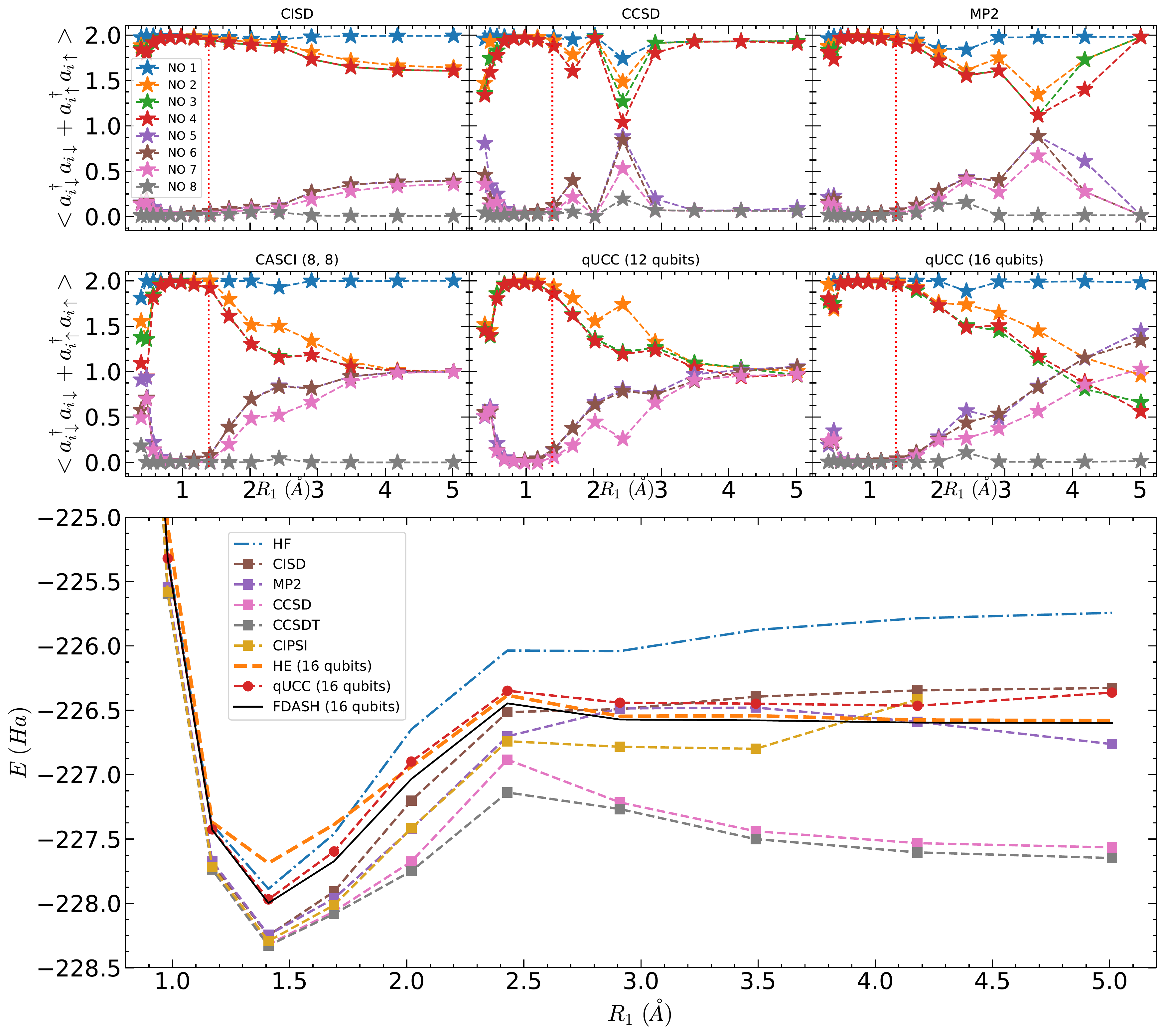}
         \end{overpic}
    \put(-455,445){(a)}
    \put(-303,445){(b)}
    \put(-151,445){(c)}
    \put(-455,338){(d)}
    \put(-303,338){(e)}
    \put(-151,338){(f)}
    \put(-455,231){(g)}
    \end{subfigure}
\caption{NOONs obtained with (a) CISD, (b) CCSD, (c) MP2, (d) CASCI, (e) and (f) qUCC methods, for distortion 1. (a) (b) and (c) were obtained through a calculation on the whole benzene system while (d) (e) and (f) were obtained within different active-space selections. They all share the same axis, and the red-dashed vertical line indicates the equilibrium geometry. Finally (g) shows the ground state energy curves obtained with different methodologies. FDASH is an abreviation for "full diagonalization of the active-space Hamiltonian".}

\label{fig:compa}
\end{figure*}

\subsection{\label{sec:level28}Comparison of quantum and classical methodologies}

In Fig. \ref{fig:compa} we compare quantum computing methodologies with state-of-the-art classical computing methods for the first distortion (similar results for distortion 2 and 3 can be found in the Appendix Fig. \ref{fig:compb}). The Natural Orbital Occupation Numbers (NOONs) are estimated from classical solvers in Fig. \ref{fig:compa} (a), (b) and (c), the 12 qubits qUCC solution in Fig. \ref{fig:compa} (e) and 16 qubits qUCC solution in Fig. \ref{fig:compa} (f). (d) shows the NOONs obtained with CASCI \cite{helgaker2014molecular} calculation made on an active-space of 8 electrons in 8 orbitals. Fig. \ref{fig:compa} (g) displays the ground state energy curve, in which the HE and qUCC results are the 16 qubits optimal guesses shown in Fig \ref{full_dist_qb}. The black curve captioned "FDASH" corresponds to the ground state energy of the 16 qubits systems obtained with a full diagonalization of the 16 qubits active-space Hamiltonians. The other energies are obtained with classical methods applied to the whole benzene system without orbital freezing.

In terms of energy (Fig. \ref{fig:compa} (e)), MP2, CCSD and CCSD(T) methods yield unphysical dissociation curves. MP2 starts to fail around $R_1 = 3.5$ \AA $\,$ and around approximately $R_1 = 2.5$ \AA $\,$ for CCSD and CCSD(T). CIPSI method has difficulties to converge starting from $R_1 = 3.5$ \AA. CISD gives a relatively low energy and a physical dissociation curve without local minima. From $R_1 = 2.5$ \AA, there is a competition between dynamic and static correlations that is not captured by all methods, especially by CCSD which is limited by its single-determinant nature. As said before, the qUCC method is very precise, especially around the equilibrium position, while the HE method gives results very close to the reference energy when the system is strongly distorted.

\begin{figure*}[t!]
     \centering
     \includegraphics[width=\textwidth]{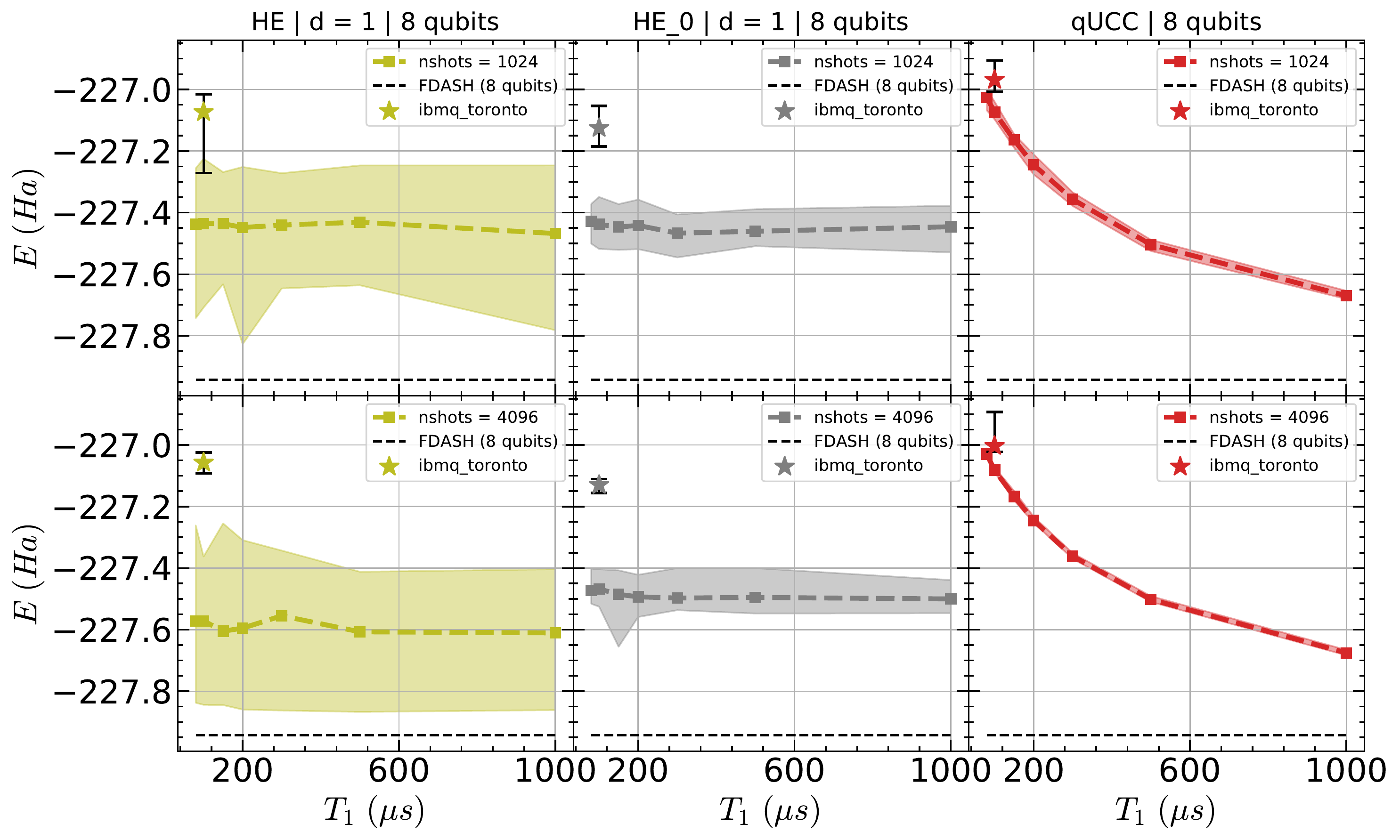}
     \caption{Ground state energy of active-space Hamiltonian of 8 qubits of benzene at $R_1 = 1.41$ \AA, as a function of $T_1$. The curves are noisy simulations while star markers were obtained with \textit{IBMQ{\_}Toronto}. The first row shows results with 1024 shots, the second one with 4096 shots. They share the same y axis. FDASH is an abreviation for "full diagonalization of the active-space Hamiltonian".}
     \label{bench_noise_T1}
\end{figure*}

The MP2 and CCSD energetic behavior can be understood with the NOONs (Fig. \ref{fig:compa} (d)). Indeed, when distortion 1 is applied to benzene, we are artificially creating 6 \chemform{C-H} pairs. Due to the symmetry of the chemical system, we expect, for extreme distortions, to avoid the interaction between these pairs and then to have non interacting orbitals with the same energy, which results in orbitals with the same occupation number. However, MP2 and CCSD NOONs fail for the same geometries in which the energies are non physical, while CISD gives relatively more realistic NOONs. Here, the CASCI NOONs acts as a reference for the qUCC NOONs. Not only the qUCC NOONs show explicitly the expected phenomena: the occupation number of NO 2-7 are gathering as $R_1$ is increasing, but also one can notice that the qUCC 12 qubits NOONs match much better the CASCI NOONs than qUCC 16 qubits results. This explains the convergence issues detailed on \ref{sec:level28} between the 12 and 16 qubits results : by taking NO 1 and 8 into consideration, the optimization cost increases unnecessarily with parameterizing orbitals that are already well described when frozen. This phenomenon coupled with a limited number of optimizations steps yields difficulties for solving larger systems on real quantum computers, even without taking into consideration the noise : At $R_1 = 1.41$ \AA, $E(CIPSI_{full}) - E(qUCC_{16 qubits}) = 0.33 \ Ha $, which means that even if active-space selection induces errors that theoretically disappear when increasing the size of the system, in practice optimization problems start inducing significant errors. In addition, it confirms the validity of the Hückel approximation as NO 1 and NO 8 remain frozen whatever the distortion.

\subsection{\label{sec:level29}Robustness against noise}

In this subsection we compare the robustness against shot and decoherence noise of HE and qUCC ansatz at the equilibrium point. Following some interesting works about the noise effects on VQE calculations \cite{fontana2021evaluating} \cite{ding2022evaluating} \cite{saib2021effect} \cite{wang2021noise}, we aim to go into more detail about some noise effects when treating such a complex system as benzene. With taking gate durations indicated in Table \ref{noisy_par} and fixing $T_2 = T_1$, we calculated the ground state energy of the 8 qubits active-space Hamiltonian benzene, for $R_1 = 1.41$ \AA, with $T_1$ varying from $80$ $\mu s$ to $1000$ $\mu s$. Fig. \ref{bench_noise_T1} shows the average of 10 trials obtained with qUCC ansatz and the average of 50 trials obtained with HE ansatz. "HE" refers to the calculations made with random initial guess, while "HE{\_0}" always had the same initial guess, which was zero. Each trial is computed with 1024 shots for the first row, 4096 shots for the second one. The dashed curves with square markers are the average of trials while the filled area represents the amplitude of variation through the trials. While the whole VQE algorithm was used for the simulations, only the last optimization step (obtained in a noiseless case) was taken and we launched it on an actual quantum computer \textit{IBMQ{\_}Toronto}, which has $T_1$ an $T_2$ approximately at $100 \ \mu s$. This was done in order to estimate the effect of noise on the estimation fo the expectation value itself as estimating the noise during the classical optimization process would run beyond the scope of this manuscript. The stars denotes the average of 10 experiments, and maximum and minimum values are represented by horizontal lines.

\begin{figure*}[t!]
     \centering
     \includegraphics[width=\textwidth]{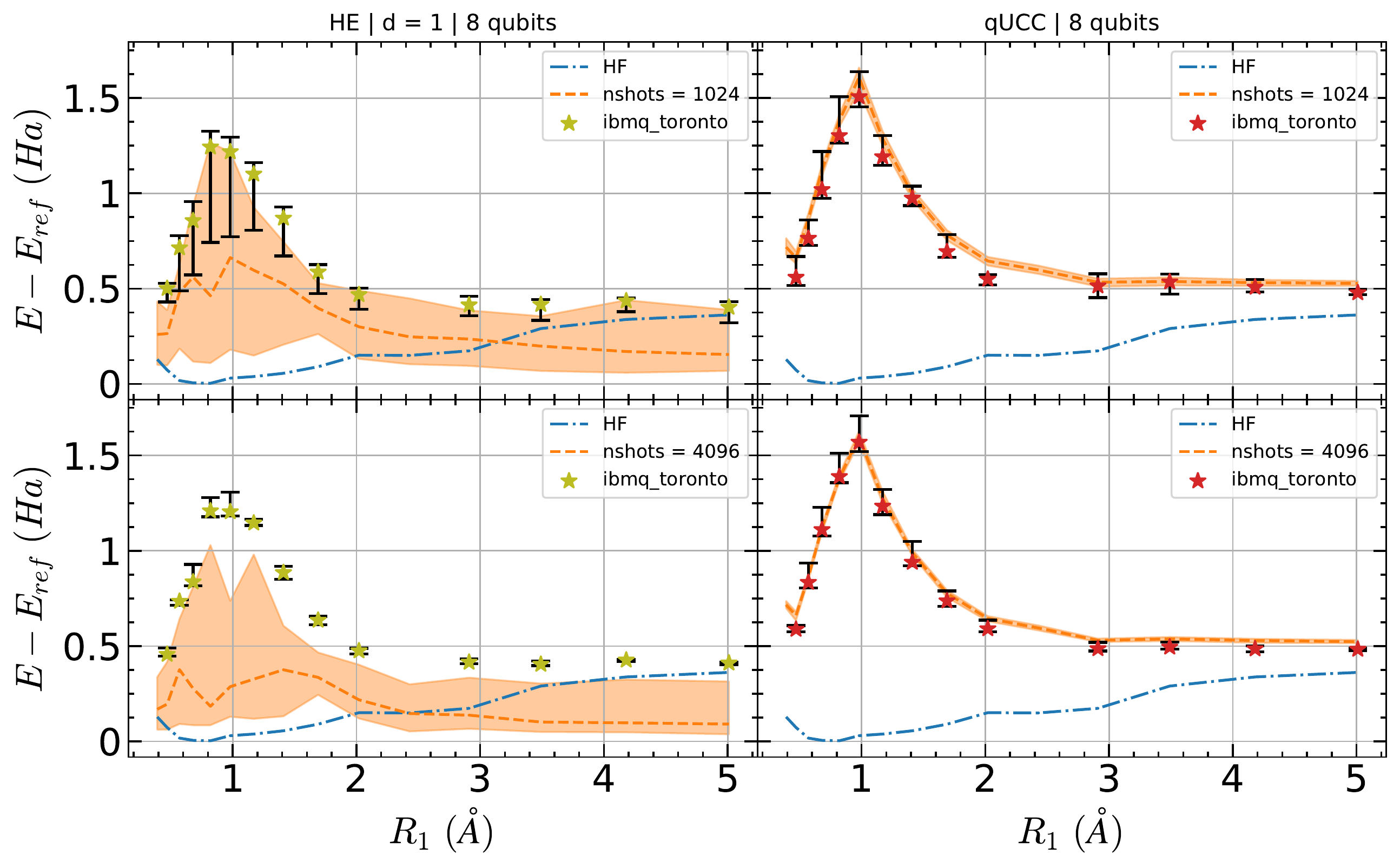}
     \caption{Difference between the ground state energies of 8 qubits system obtained with HE and qUCC ansatz, either with our noise model or real \textit{IBMQ{\_}Toronto}, and the reference obtained with full diagonalization of 8 qubits system, for the first distortion. The first row shows results with 1024 shots, the second one with 4096 shots. All figures share the same y axis.}
     \label{bench_noise_dist1}
\end{figure*}

This figure shows that increasing the number of shots does not have obvious consequences, and it confirms some claims made in Section \ref{sec:level25}. First, one can see that from 1024 to 4096 shots the variance of qUCC results decreases while the average remains constant. This confirms that qUCC is much more affected by decoherence noise than measurement noise. Furthermore, for HE results it is the opposite : the average has dropped by 0.2 Ha, independently of T1, while the variance of HE results has the same order of magnitude. This confirms the greater sensibility of HE to shot noise than hardware noise. On the other hand, for "HE{\_}0" results, neither the average nor the variance changed when increasing the number of shots. This seems to illustrate that some areas of the optimization landscape are robust against noise - initializing a state in $|0\rangle\otimes ...|0\rangle$ makes it more immune against qubit relaxation than any string involving $|1\rangle$ would be. In addition, the difference between extreme values is 0.34 Ha for HE without noise (Fig.5, first row first column), 0.46 Ha for HE with noise and 0.15 Ha for HE\_0 with noise. This shows that, for HE ansatz, the choice of a random initial guess has an higher impact on the convergence than the noise itself. On the other hand, an extrapolation on qUCC results yields $T_1 \geq 36 \ ms$ to reach a $10^{-2}$ precision, or $T_1 \geq 3.6 \ s$ for being close to be within $10^{-4}$ of the reference. It gives a numerical confirmation that the qUCC ansatz will never be compatible with actual NISQ hardwares without noise mitigation.

Regarding real QPU calculations, even if qUCC results are quite close to the noise model both quantitatively and qualitatively, HE results a bit further away from it despite the slight overlap between HE error bar and filled area. However, our noise model is a simplified and the values for $T_1$ and $T_2$ change dynamically between calibrations of IBMQ hardware. Furthermore, the connectivity on IBMQ hardware is not all-to-all and this is compensated with additional SWAP gates.

In Fig. \ref{bench_noise_dist1}, we compare the results obtained with our noise model (see Table \ref{noisy_par}) with real QPU calculations thanks to \textit{IBMQ{\_Toronto}}, for 8 qubits active-space Hamiltonians of benzene under distortion 1. As before, the VQE has been fully used in our noise model case, and only the last optimization step (yield by a noiseless computation loop) was processed on quantum computer. The dashed curve corresponds to the noise model and holds an average of 50 trials, while filled area illustrates the amplitude of variation of the results. The stars corresponds to an average of 10 real QPU experiments, with maximum and minimum values represented by horizontal dashes. In both case, the reference is the ground state energy of the 8 qubits system, obtained by a full diagonalization of the active-space Hamiltonians of 8 qubits. 

First, one can see that the average of qUCC experiments match quite well our noise model, while the average of HE experiments are $< 0.3$ Ha away from it. However, despite the slight quantitative gap between HE ansatz results on QPUs and HPCs, the error bars are close to the value obtained for the noise model. One can also notice that the curves have the same shape, meaning that the influence of noise with $R_1$ is similar in both situations. Especially, the noise is much more dominating around the equilibrium position than for extreme distortions, which confirms the previous statements. This can also be seen on 16 qubits results (see Fig.\ref{bench_noise_dist1_16qb} in Appendix \ref{appendix_curves}). Moreover, increasing the number of shots has no significant impact on the average of experiments but it decreases the variance of results. In a more chemical point of view, even for extreme distortions, actual QPUs with HE ansatz would not enable to outperform HF calculations.

\section{Conclusion}
\label{sec_conc}

In conclusion we have estimated the feasibility to execute families of hardware efficient quantum computing ansatze and quantum unitary coupled cluster ansatze on near term quantum computers. By incorporating a realistic noise model we find that hardware efficient ansatze could be executed on near term hardware, giving a better precision than mean-field methods far away from the charge equilibrium point. On the other hand side qUCC is superior to mean-field methods but will remain a method for simulators in the pre-error-correction era. The qUCC method also preserves well the particle number of individual orbitals. Noise remains a central issue, particularly around the equilibrium position. For all approaches, noise appears consequent but its importance is shown to be more critical in the case of the qUCC ansatz that could be a non-operational method on the NISQ era real quantum computers for such a large system as benzene. Real experiments on IBMQ show that our noise model is describing well the noisy operation of a QPU both qualitatively and somewhat quantitatively.
Overall, more research is needed to develop advanced noise-resistant ansatze, especially for those based on qUCC. Regarding the fact that qUCC is a lot more sensitive to the hardware noise than to the shot noise and knowing that reaching an unlimited number of measurements seems difficult on a real machine, it seems important to work out noise-reduction techniques in this particular area. In addition, ansatze continue to evolve and recent works directed toward Variational Quantum Eigensolvers (VQE) may improve the robustness of qUCC-based methodologies in a near future \cite{bharti2022noisy, tilly2021variational}. Anecdotally, although Adapt-VQE methodologies are state-of-the-art on the algorithmic side for QPUs, the number of gradients to be calculated remains a huge problem for practical implementation \cite{haidar2022open}, due to large number of gradients evaluations that scale at least suboptimally with system size and would overconsume actual QPU ressources. Making a calculation on real QPU requires a compromise between circuit depth and number of executions. Thus we foresee that if Adapt-VQE-like algorithms would reduce circuit depth, they will also enhance the complexity of each execution. In this form, it would have limited impact on large systems, especially on NISQ era, as system size is not only limited by the noise but also by the optimizations issues that strongly increases with it. Therefore, our future path would be directed towards solving this problematic.

Our findings further suggest that the HE ansatz is sensitive to the random initial guess of the parameters, commonly leading to false minima of the optimizer. Furthermore, increasing the size of the active space can lead to a similar result - a solution which is trapped in a local minimum. Finally, in contrast to classical methods, quantum computing methodologies manage to preserve physically relevant orbital occupancies.

\begin{acknowledgments}
We wish to acknowledge Mohammad Haidar for the classical simulations, and Jan Reiner and Arseny Kovyrshin for useful discussions. W.S. and M.R. acknowledge funding from European Union's Horizon 2020 research and innovation programme, more specifically the $\langle$NE$|$AS$|$QC$\rangle$ project under grant agreement No. 951821.
\end{acknowledgments}

\appendix

\section{Second quantization Hamiltonian}

Generally, a molecule is composed of atoms, themselves composed of nuclei and electrons. The Hamiltonian of any molecule can be written as:
\begin{equation}
H = T + V,
\end{equation}
with $T$ as the kinetic energy operator and $V$ the potential energy operator. Using atomic units, $T$ describes the movement of the particles and $V$ describes the Coulomb interaction between them :

\begin{equation}
T = \underbrace{-{\sum_{i=1}^{N} \frac{1}{2}\nabla_{i}^{2}}}_{\text{electon}} - \underbrace{- {\sum_{A=1}^{M} \frac{1}{2}\nabla_{A}^{2}}}_{\text{nuclei}}
\end{equation}
\begin{equation}
V = \underbrace{- \sum_{i,A} \frac{Z_A}{r_{iA}}}_{\text{electron-nuclei}} + \underbrace{\sum_{j>i} \frac{1}{r_{ij}}}_{\text{electron-electron}} + \underbrace{\sum_{B>A} \frac{Z_AZ_B}{r_{AB}}}_{\text{nuclei-nuclei}}
\end{equation}

These equations are simplified when you work with the Born-Oppenheimer approximations, by neglecting the kinetic energy term of nuclei and the Coulomb interaction term between them. So, the Hamiltonian is:

\begin{equation}
H = -{\sum_{i} \frac{1}{2}\nabla_{i}^{2}} - \sum_{i,A} \frac{Z_A}{r_{iA}} +  \sum_{j>i} \frac{1}{r_{ij}}.
\end{equation}

The principle of second quantization consists on rewriting the Hamiltonian with the creation and annihilation operators, $a^\dagger$ and $a$, that acts on the occupation number vector:

\begin{equation}
H = \sum_{p,q} h_{pq}a_p^\dagger a_q + \sum_{p,q,r,s}\frac{h_{pqrs}}{2}a_p^\dagger a_q^\dagger a_r a_s
\end{equation}

The first term describes both the kinetic energy and the Coulomb interaction with nuclei of an electron, while the second term describes the two-body Coulomb interaction between pairs of electrons. With $\mathbf{x} = (\mathbf{r},\sigma)$, these terms can be written as :

\begin{equation}
\forall p,q, \ \ h_{pq} = \int \mathrm{d}\mathbf{x} \phi_{p}^{*}(\mathbf{x})\left(-\frac{1}{2}\nabla^2 - \sum_{A} \frac{Z_A}{|\mathbf{r}-R_A|}\right) \phi_{q}(\mathbf{x}),
\end{equation}

\begin{eqnarray}
\forall p,q,r,s, \ \ h_{pqrs} &=& \int \mathrm{d}\mathbf{x}_1 \mathrm{d}\mathbf{x}_2 \phi_{p}^{*}(\mathbf{x}_1)\phi_{q}^{*}(\mathbf{x}_2) \nonumber \\ & & \times \left(\frac{1}{|\mathrm{r}_1-\mathrm{r}_2|}\right) \phi_{r}(\mathbf{x}_1)\phi_{s}(\mathbf{x}_2).
\end{eqnarray}

Each $\phi(\mathbf{x}) = \phi(\mathbf{r},s)$ represents a spin-orbital of the basis set. Finally, the more elaborate the basis set, the larger is the Hamiltonian : if we stick with a basis with $N_{orb}$ orbitals, then it has $2N_{orb}$ spin-orbitals, which results in a full Hamiltonian stored in a $(2^{2N_{orb}},2^{2N_{orb}})$ matrix.

\section{The Jordan-Wigner transformation}
\label{appendix_JW}
Quantum computers are composed out of qubits, whose state is modified by rotations in Bloch sphere, which are performed by single qubit quantum gates. In order to be understandable by quantum computers, the second quantized Hamiltonian has to be transformed into a new form - obeying the fermionic algebra of electrons but also behaving to the SU(2) group behaviour of qubits. While the occupation number vector is quite easy to transform, by asserting a qubit to $\ket{0}$ if a spin-orbital is empty, and $\ket{1}$ if a spin-orbital contains an electron, the main problem is the conservation of the antisymmetric properties of the system. The Jordan-Wigner transformation converts the creation and annihilation operators, while preserving their antisymmetric properties \cite{mcardle2020quantum}:

\begin{equation}
\forall i < N_{qubits}, \ \ a_i^\dagger \rightarrow \left(\prod_{j<i}\sigma_j^{Z}\right)\frac{\sigma_{i}^{X} - i\sigma_{i}^{Y}}{2}
\end{equation}
\begin{equation}
\forall i < N_{qubits}, \ \ a_i \rightarrow \left(\prod_{j<i}\sigma_j^{Z}\right)\frac{\sigma_{i}^{X} + i\sigma_{i}^{Y}}{2}
\end{equation}

Finally, the Hamiltonian of the system can be written as a sum of Pauli strings \cite{moll2018quantum}:

\begin{equation}
H_{JW} = \sum_{j} h_j P_j
\end{equation}

\section{Idle noise model}
\label{appendix_idle_noise}

Briefly, let us consider a qubit in the state $\ket{\phi} = \alpha \ket{0} + \beta \ket{1}$. The density matrix $\rho_{0}$ of the pure state is :

\begin{equation}
    \rho_{0} = \ket{\phi}\bra{\phi} = 
    \begin{pmatrix}
    |\alpha|^2 & \alpha \beta^* \\
    \alpha^*\beta & |\beta|^2 \\
    \end{pmatrix}.
\end{equation}

The Kraus operators for pure dephasing \cite{bogdanov2013modeling} are, with $p_{ph}(t)$ being the decaying law of the phase :

\begin{equation}
    E_{0} = 
    \begin{pmatrix}
    1 & 0 \\
    0 & \sqrt{1-p_{ph}(t)} \\
    \end{pmatrix} \ \ \ \ \text{and} \ \ \ \
    E_{1} = 
    \begin{pmatrix}
    0 & 0 \\
    0 & \sqrt{p_{ph}(t)} \\
    \end{pmatrix}
\end{equation}

Then the density matrix which takes into account this phenomenon becomes : 

\begin{equation}
    \rho = \sum_{k} E_k \rho_{0} E_k^\dagger = 
    \begin{pmatrix}
    |\alpha|^2 & \alpha \beta^*\sqrt{1-p_{ph}(t)} \\
    \alpha^*\beta\sqrt{1-p_{ph}(t)} & |\beta|^2 \\
    \end{pmatrix}
\end{equation}

Moreover, the Kraus operators for amplitude damping are, with $p_{a}(t)$ the decreasing law of the amplitude :

\begin{equation}
    E_{0} = 
    \begin{pmatrix}
    1 & 0 \\
    0 & \sqrt{1-p_{a}(t)} \\
    \end{pmatrix} \ \ \ \ \text{and} \ \ \ \
    E_{1} = 
    \begin{pmatrix}
    0 & \sqrt{p_{a}(t)} \\
    0 & 0 \\
    \end{pmatrix}
\end{equation}

Then the density matrix which takes into account this phenomenon becomes :

\begin{equation}
    \rho = 
    \begin{pmatrix}
    |\alpha|^2+ |\beta|^2 p_{a}(t) & \alpha \beta^*\sqrt{1-p_{ph}(t)} \\
    \alpha^*\beta\sqrt{1-p_{ph}(t)} & |\beta|^2(1 - p_{a}(t)) \\
    \end{pmatrix}
\end{equation}
At the end, one remarks that the pure dephasing impacts only the off-diagonal terms. Therefore, we can rewrite the density matrix for the two combined phenomenon, and considering time-exponential relaxations $p_{a}(t) = 1 - e^{-\frac{t}{T_1}}$ and $p_{ph}(t) = 1 - e^{-\frac{2t}{T_{2}}}$ :

\begin{equation}
    \rho =
    \begin{pmatrix}
    |\alpha|^2+ |\beta|^2(1 - e^{-\frac{t}{T_1}}) & \alpha \beta^*e^{-\frac{t}{T_{2}^{'}}} \\
    \alpha^*\beta e^{-\frac{t}{T_{2}^{'}}} & |\beta|^2 e^{-\frac{t}{T_1}} \\
    \end{pmatrix}
\end{equation}

with $T_2^{'}$ defined as $\frac{1}{T_2^{'}} = \frac{1}{T_2} + \frac{1}{2T_1}$. Generalized to the whole qubit register, one is able to incorporate the effect of noise in the simulations.

\section{Supplementary results}
\label{appendix_curves}

\begin{figure*}[t!]
     \centering
     \begin{subfigure}[H]{\textwidth}
         \centering
         \begin{overpic}[width=\textwidth]{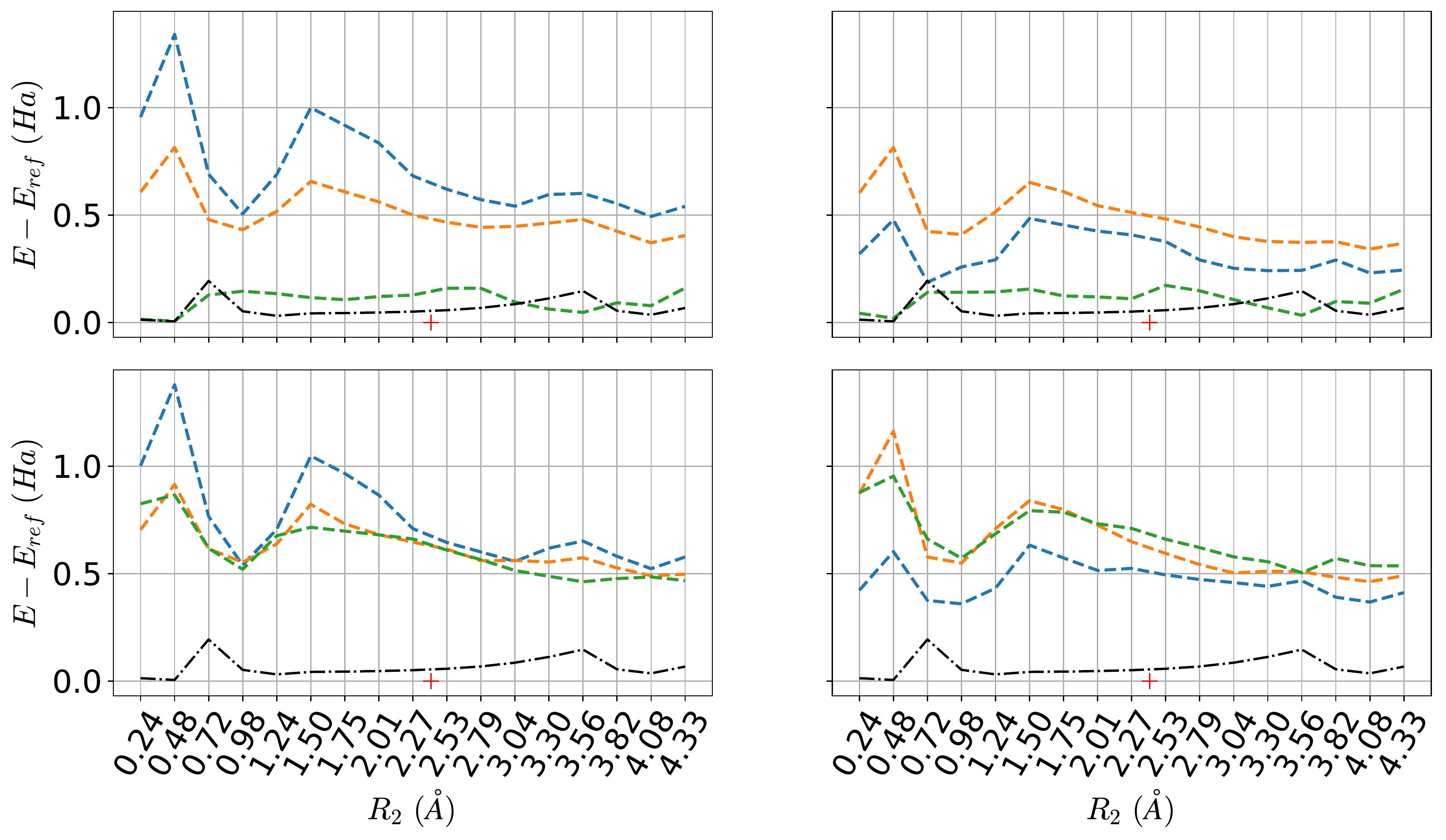}
    \put(26,58){$d=1$}
	\put(75,58){$d=2$}
	\put(-5,58){(a)}
	\put(-5,0){(b)}
    \end{overpic}	
         \label{compHEdist2}
     \end{subfigure}
     \hfill
     \begin{subfigure}[H]{\textwidth}
         \centering
         \includegraphics[width=\textwidth]{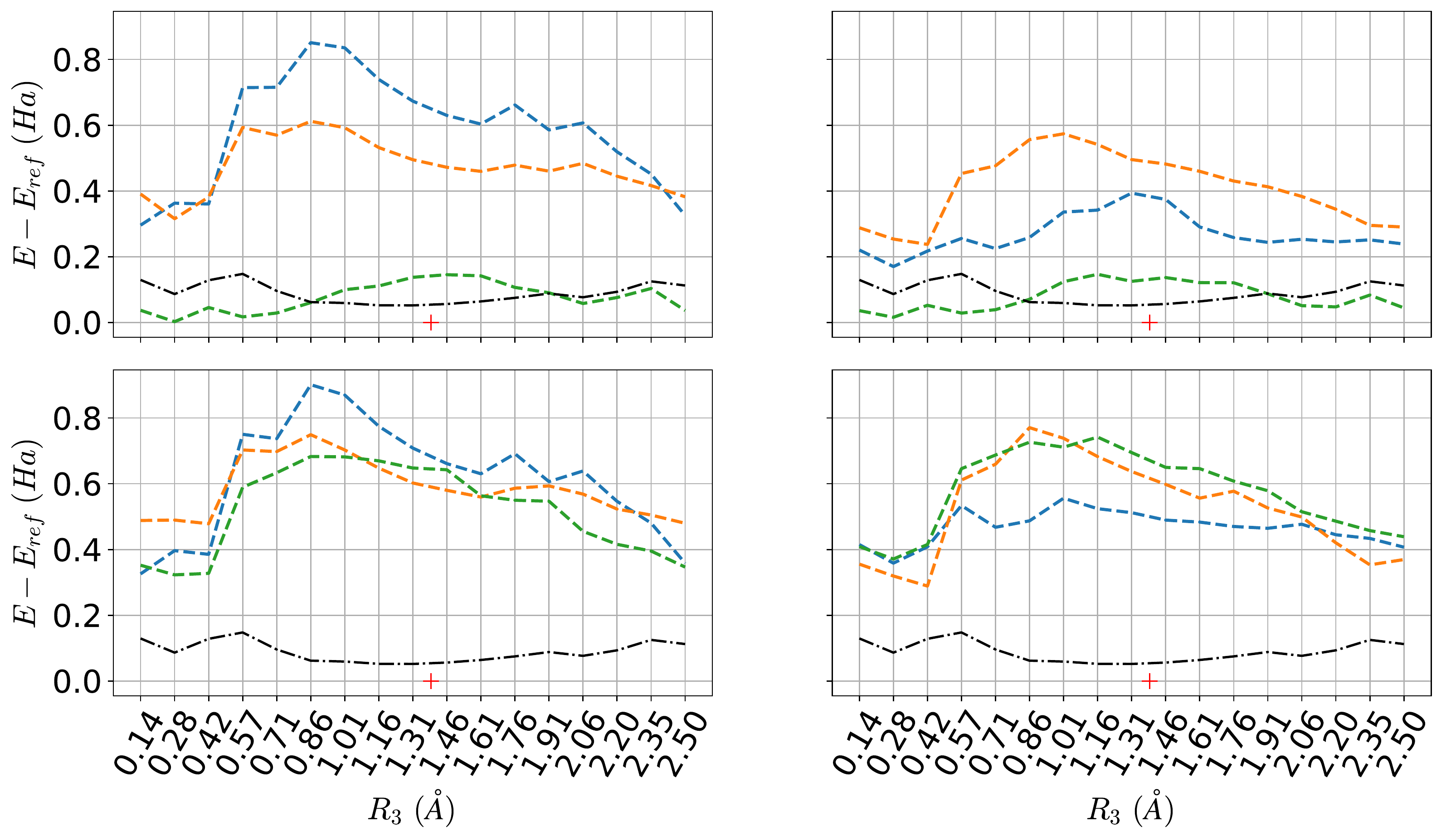}
         \label{compHEdist3}
     \end{subfigure}
    \caption{Difference between the ground state energy curves of $8$ qubits system obtained with HE ansatz for (a) the second distortion and (b) the third distortion, and the reference obtained with full diagonalization of $8$ qubits system.}
    \label{compHEdist_full_appendix}
\end{figure*}

\begin{figure*}[t!]
     \begin{subfigure}[b]{\textwidth}
         \centering
         \includegraphics[width=\textwidth]{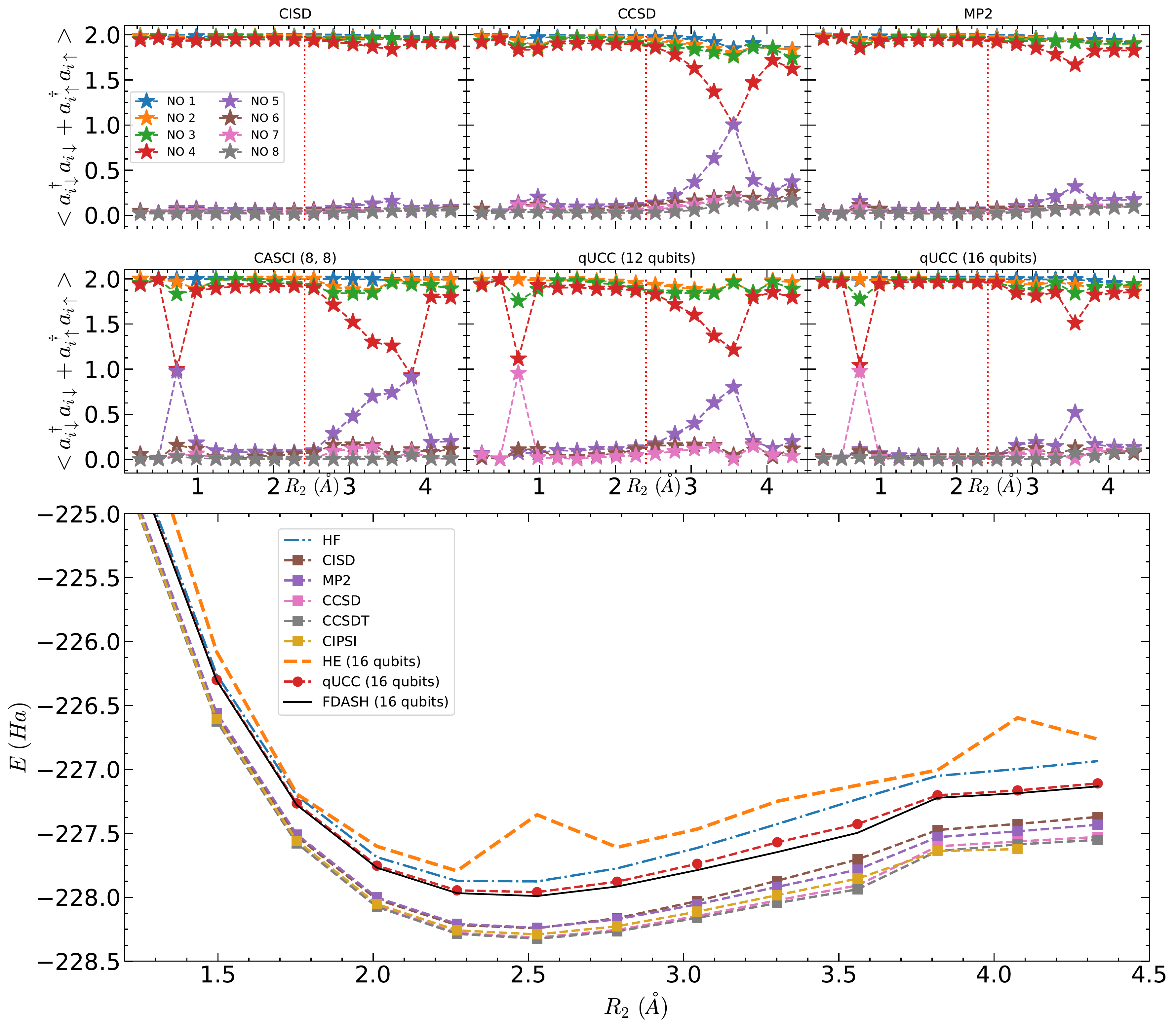}
     \end{subfigure}
     \put(-510,445){(a)}
     \put(-333,445){(b)} 
     \put(-166,445){(c)}
     \put(-510,290){(d)}
     \put(-510,142){(e)}  
     \end{figure*}
     \begin{figure*}[t!]\ContinuedFloat
     \begin{subfigure}[b]{\textwidth}
         \centering
         \includegraphics[width=\textwidth]{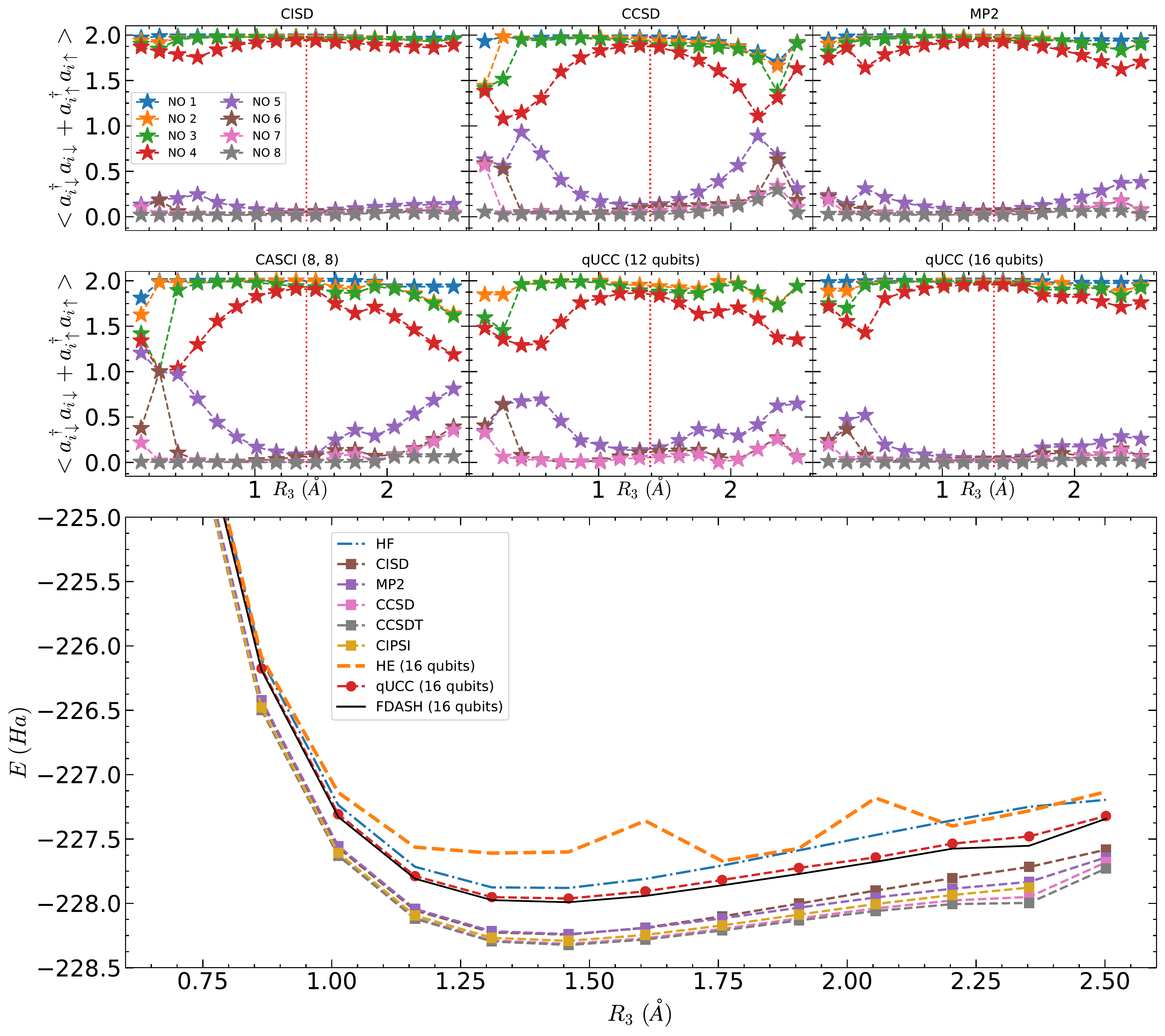}
     \end{subfigure}
     \put(-510,445){(f)}
     \put(-333,445){(g)} 
     \put(-166,445){(h)}
     \put(-510,290){(i)}
     \put(-510,142){(j)}  
     \caption{NOONs obtained with (a, h) CISD, (b, i) CCSD, (c, j) MP2, (d, k) CASCI, (e, l) and (f, m) qUCC methods, for distortion 2 and 3. (a, h) (b, i) and (c, j) were obtained through a calculation on the whole benzene system while (d, k) (e, l) and (f, m) were obtained within different active-space selections. They all share the same axis, and the red-dashed vertical line indicates the equilibrium geometry. Finally (g, n) show the ground state energy curves obtained with different methodologies. FDASH is an abreviation for "full diagonalization of the active-space Hamiltonian".}
     \label{fig:compb}
\end{figure*}

\begin{figure*}[t!]
     \begin{subfigure}[b]{\textwidth}
         \centering
         \includegraphics[width=0.78\textwidth]{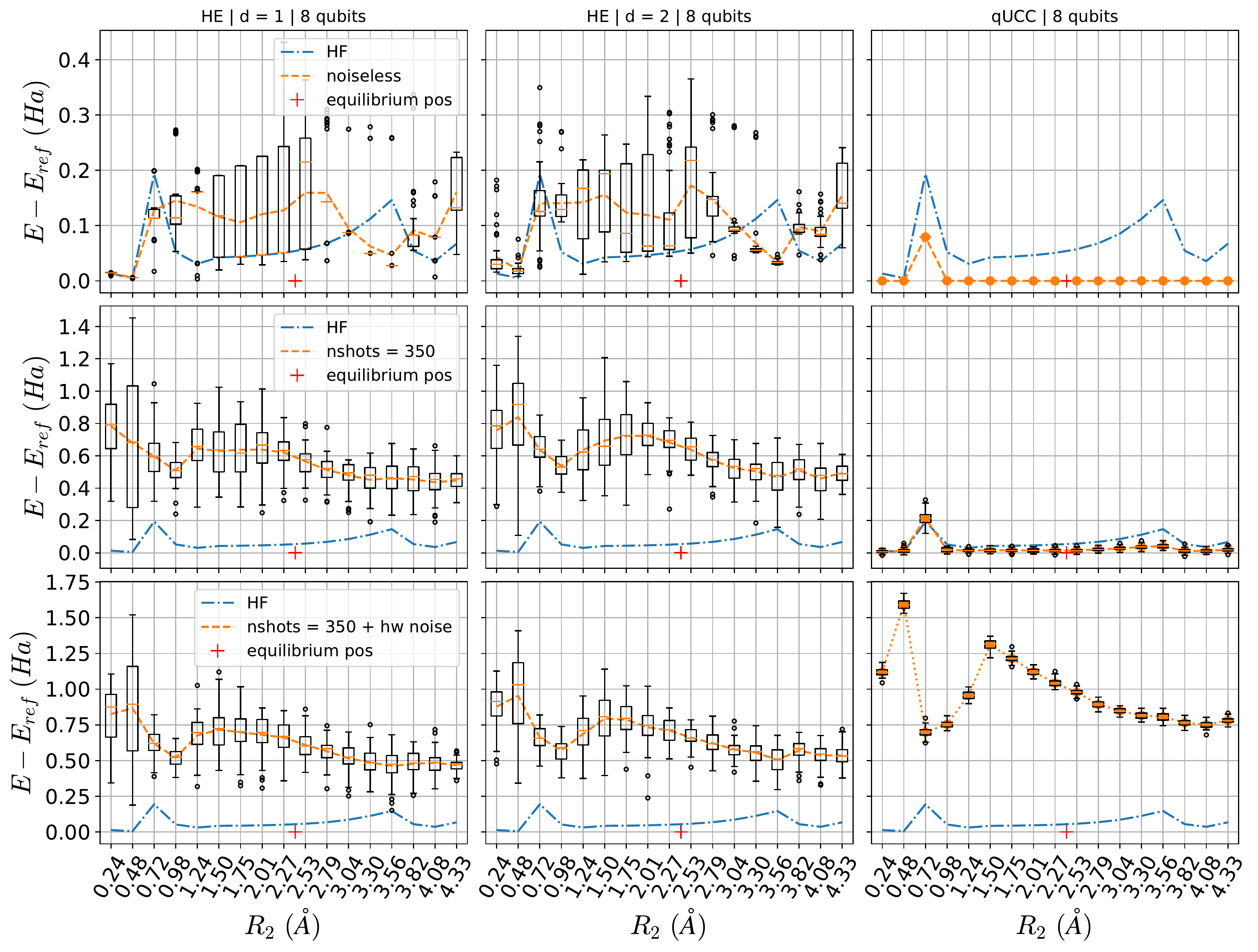}
         \label{9sub_dist2}
     \end{subfigure}
     \put(-510,290){(a)}
     \end{figure*}
     \begin{figure*}[t!]\ContinuedFloat
     \begin{subfigure}[b]{\textwidth}
         \centering
         \includegraphics[width=0.78\textwidth]{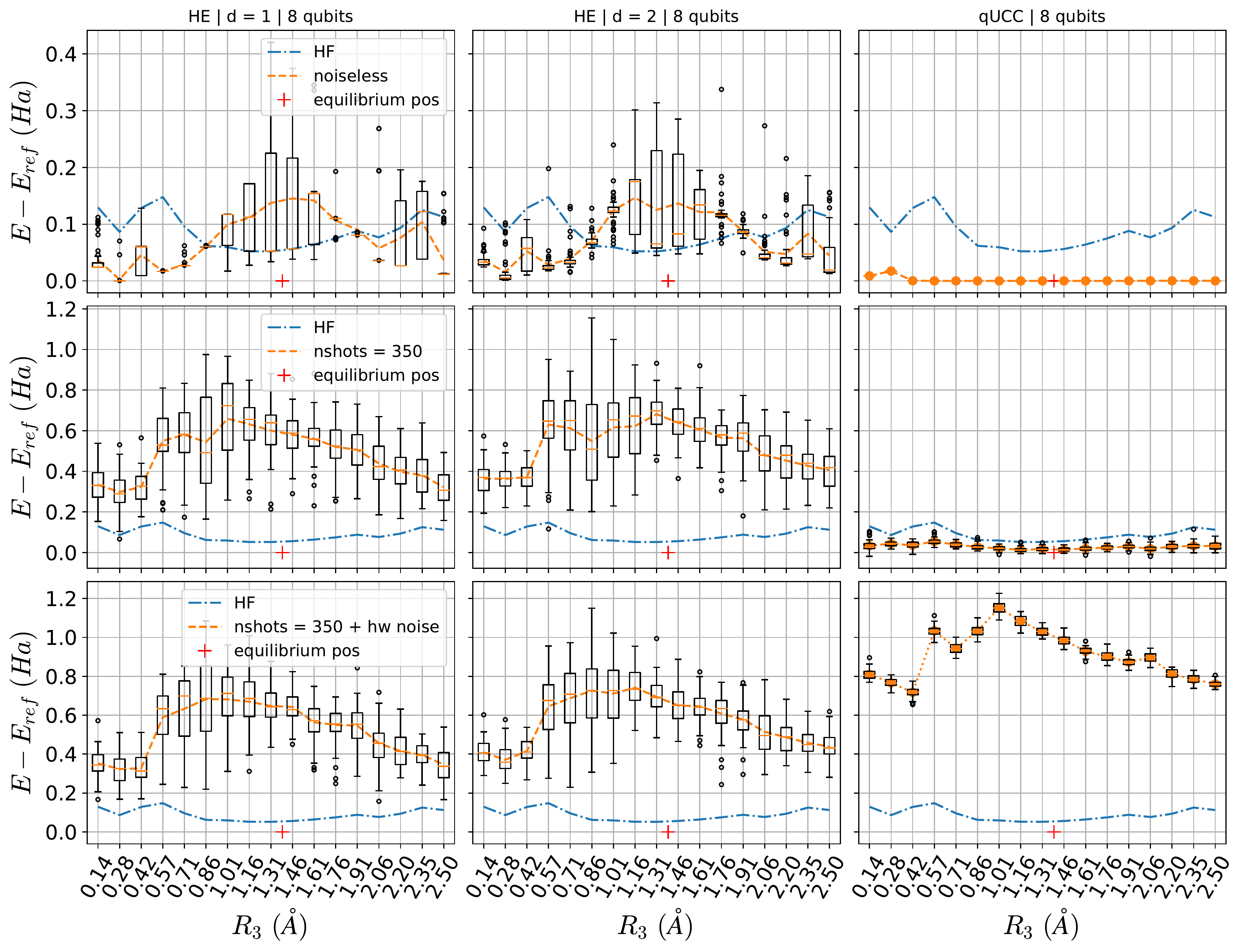}
         \label{9sub_dist3}
     \end{subfigure}
     \put(-510,290){(b)}
     \caption{Difference between the ground state energies of 8 qubits system obtained with HE and qUCC ansatz for (a) the second distortion and (b) the third distortion, and the reference obtained with full diagonalization of 8 qubits system.}
     \label{9_sub_full_appendix}
\end{figure*}

\begin{figure*}[t!]
     \begin{subfigure}[b]{\textwidth}
         \centering
         \includegraphics[width=0.8\textwidth]{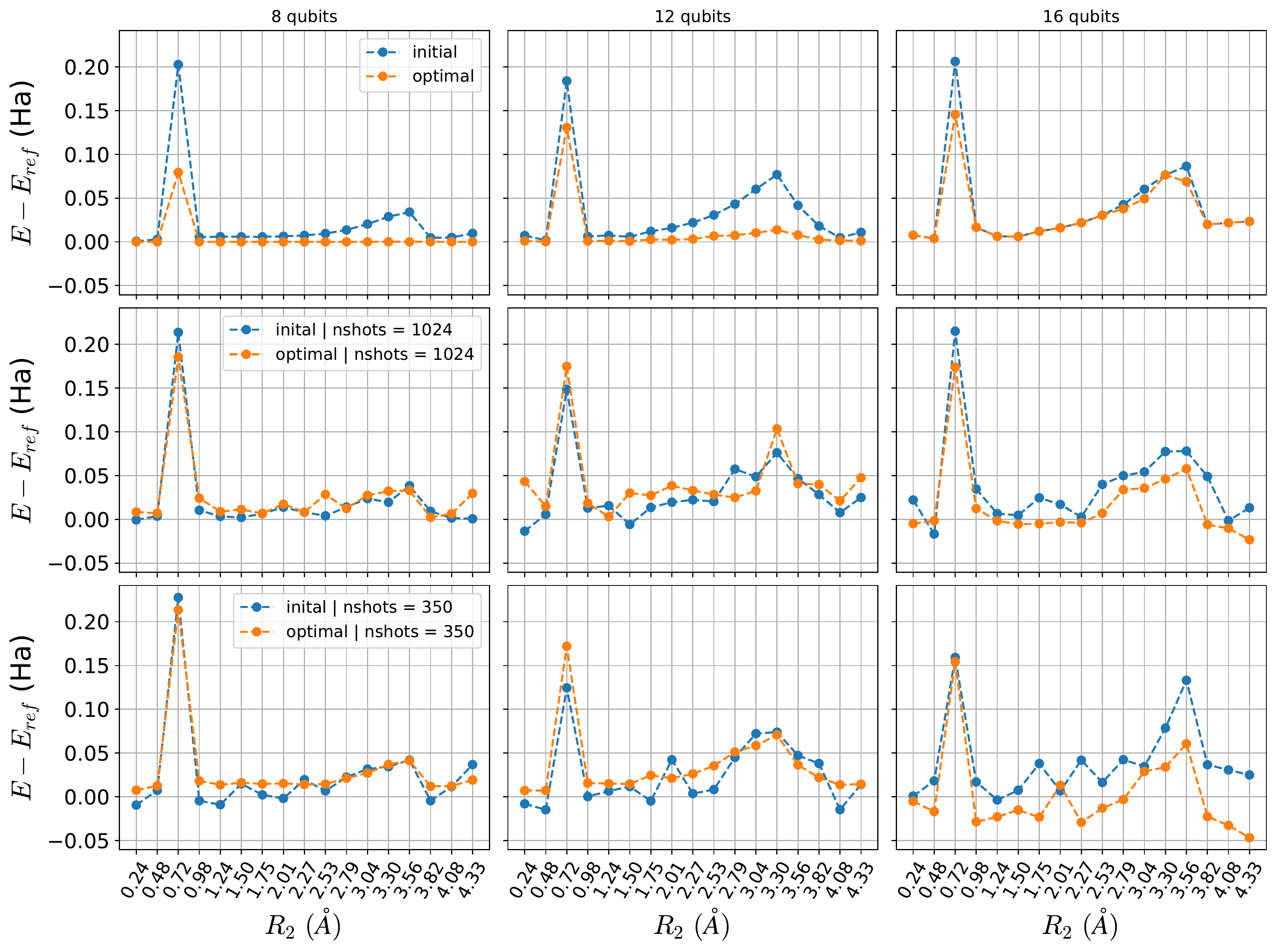}
         \label{UCC_init_dist2}
     \end{subfigure}
     \put(-510,290){(a)}
     \end{figure*}\begin{figure*}[t!]\ContinuedFloat
     \begin{subfigure}[b]{\textwidth}
         \centering
         \includegraphics[width=0.8\textwidth]{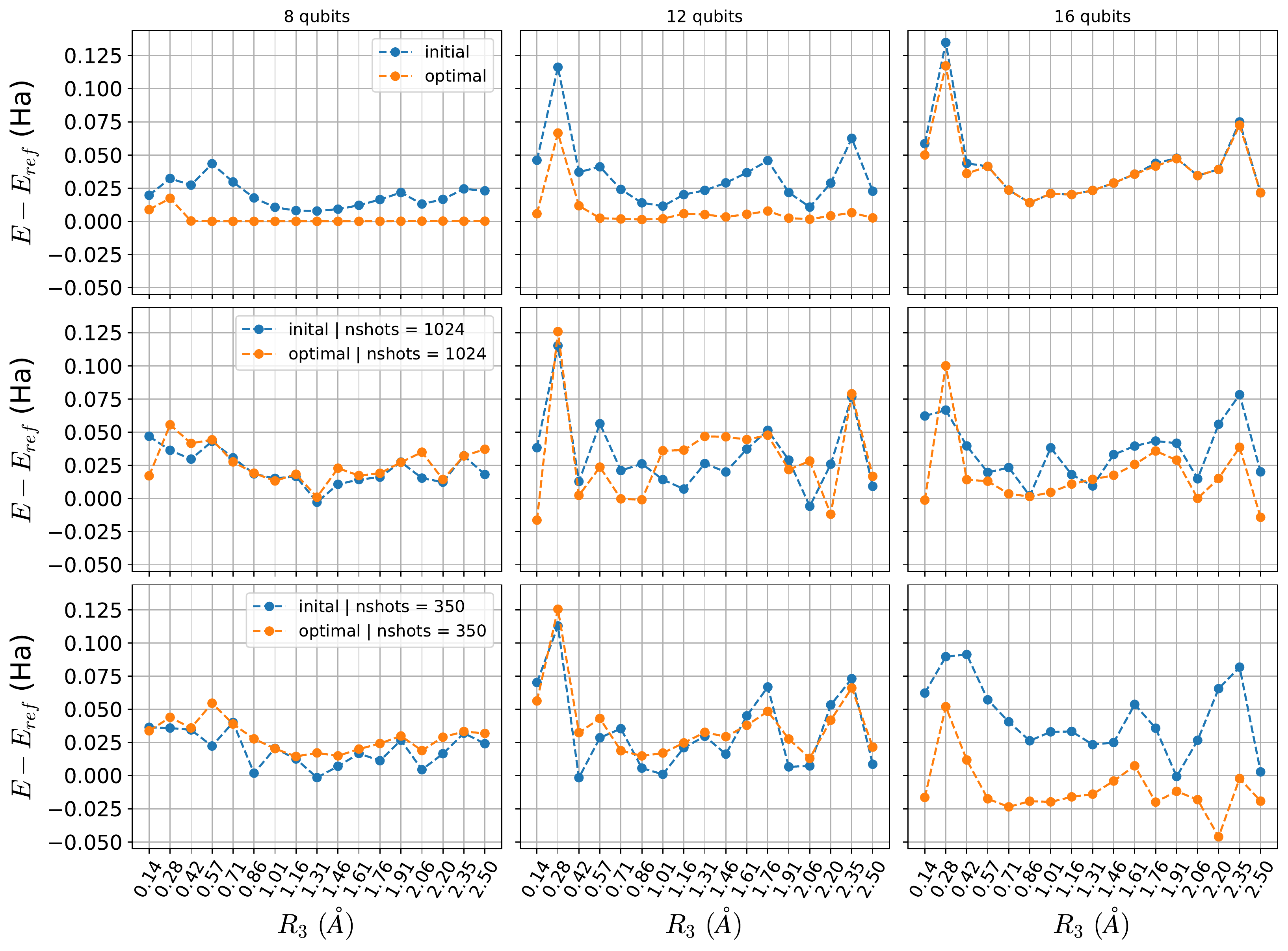}
         \label{UCC_init_dist3}
     \end{subfigure}
     \put(-510,290){(b)}
    \caption{Comparison between initial and optimal guess with qUCC ansatz, for (a) the second distortion and (b) the third distortion, with $E_{ref}$ the energies obtained with full diagonalization of the active-space Hamiltonian, with corresponding number of qubits written on top of figures.}
    \label{UCC_init_full}
\end{figure*}

\begin{figure*}[t!]
     \begin{subfigure}[b]{\textwidth}
         \centering
         \includegraphics[width=0.9\textwidth]{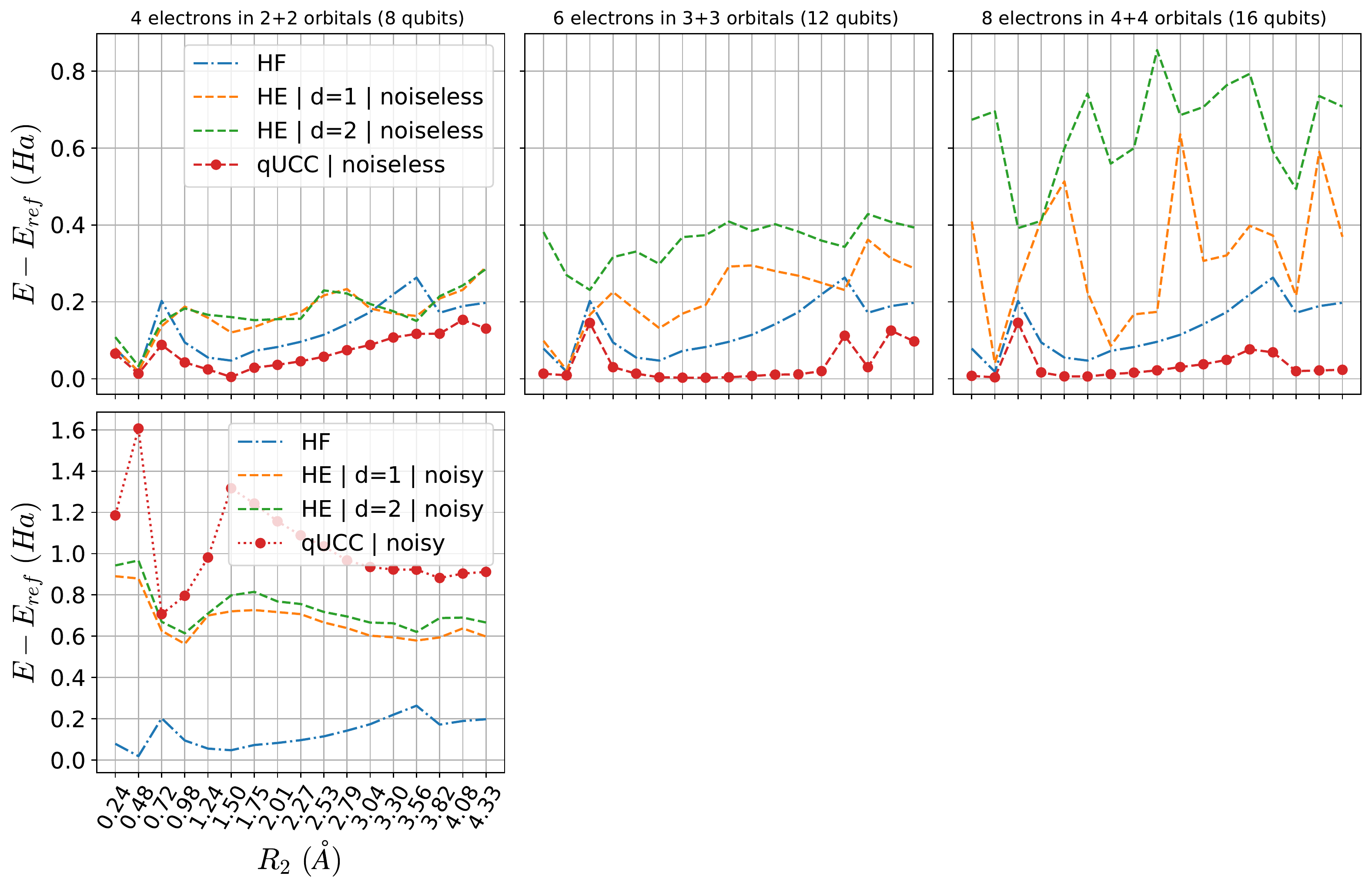}
         \label{full_dist2}
     \end{subfigure}
	\put(-525,275){(a)}
     \end{figure*}
     \begin{figure*}[t!]\ContinuedFloat
     \begin{subfigure}[b]{\textwidth}
         \centering
         \includegraphics[width=0.9\textwidth]{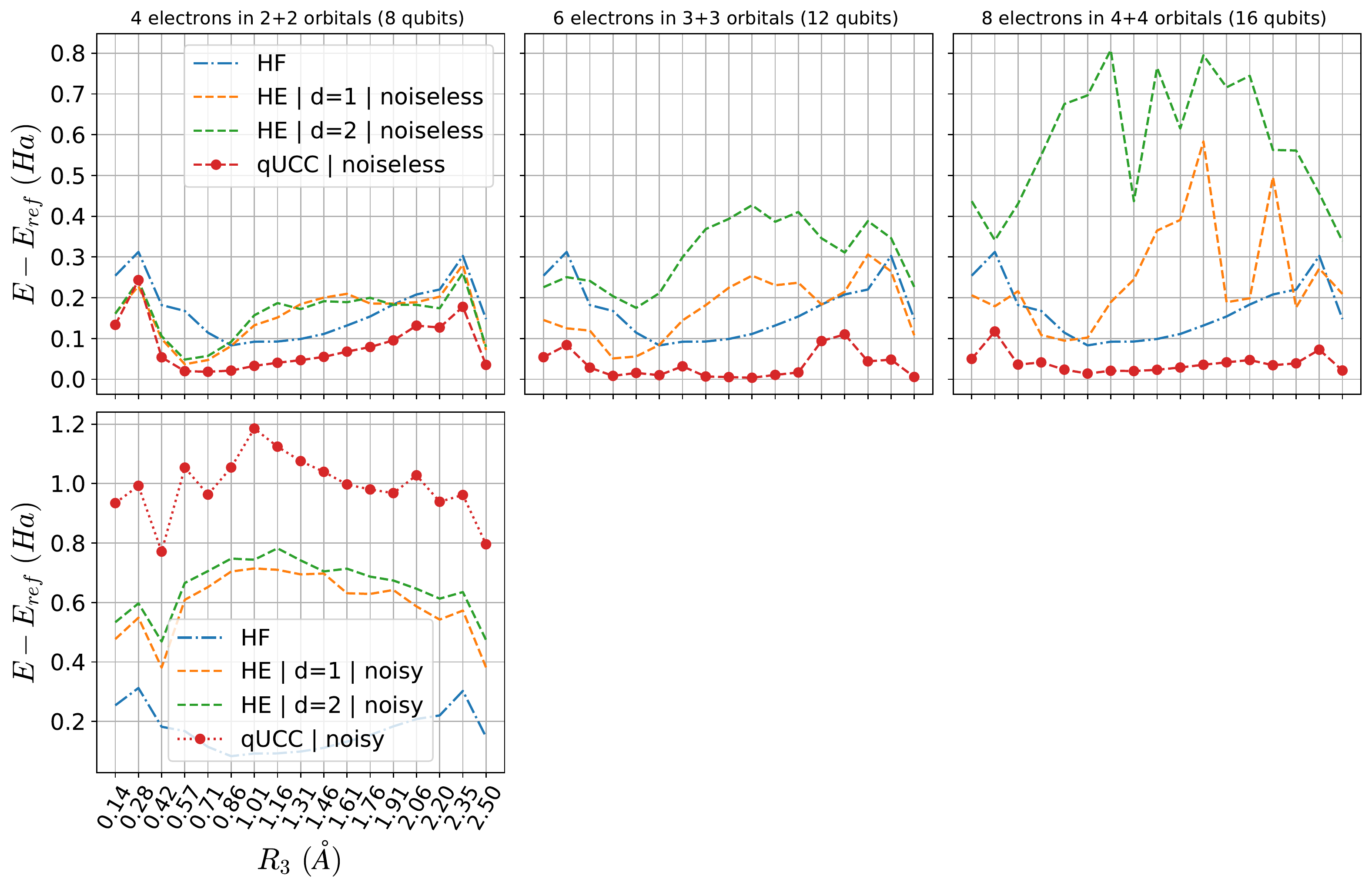}
         \label{full_dist3}
     \end{subfigure}
     \put(-525,275){(b)}
    \caption{Comparison of ground state energy curves of different number of qubits, for (a) the second distortion and (b) the third distortion, with the reference energy being obtained with full diagonalization of the active-space Hamiltonian, with corresponding number of qubits written on top of figures.}
    \label{full_dist_qb_B}
\end{figure*}

\begin{figure*}[t!]
     \centering
     \includegraphics[width=\textwidth]{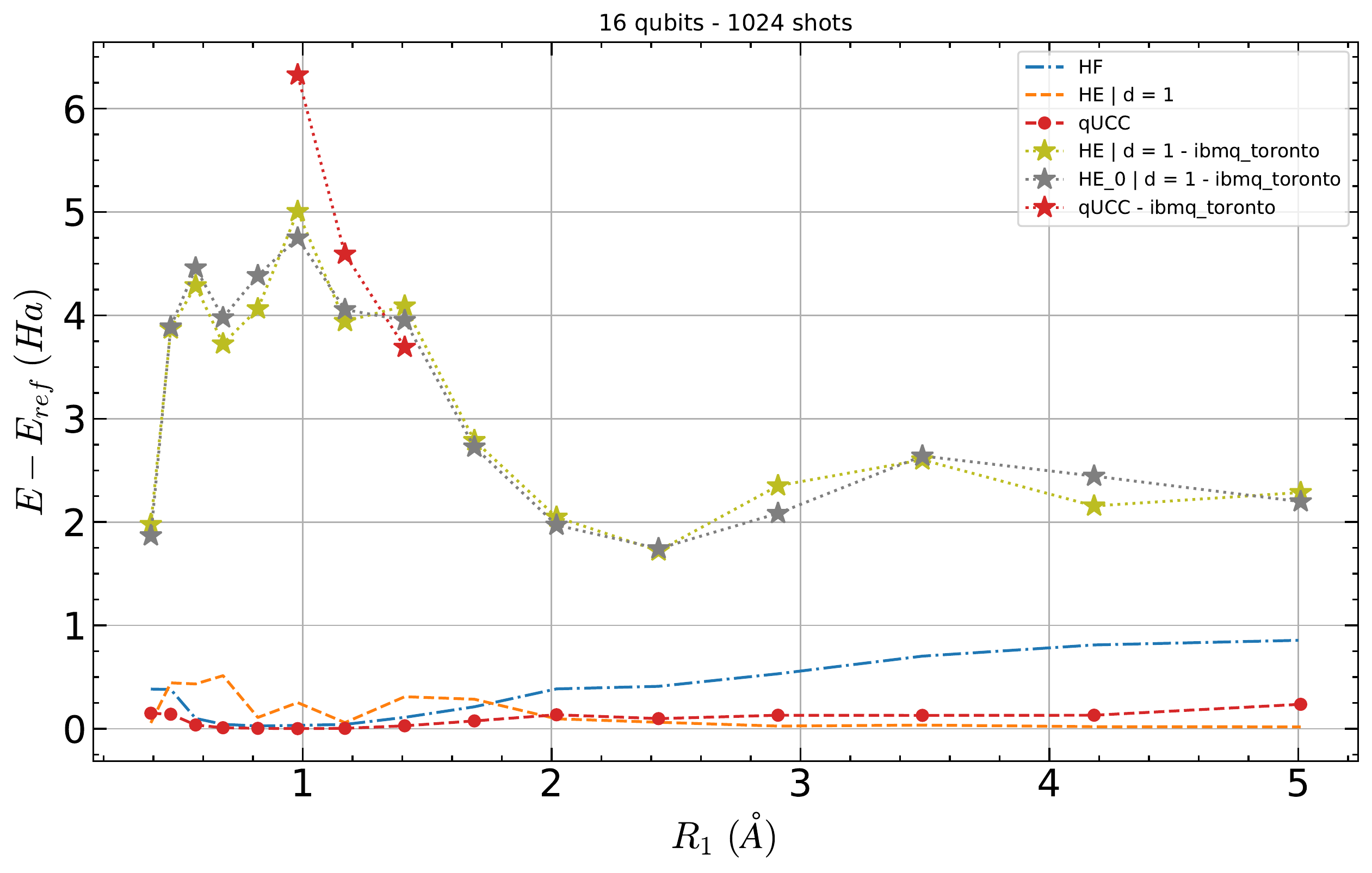}
     \caption{Difference between the ground state energies of 16 qubits systems obtained with HE and qUCC ansatz, for the first distortion. Dashed lines are obtained with noiseless simulations on QLM while real \textit{IBMQ{\_}Toronto} experiments are shown with stars. The reference is obtained with full diagonalization of 16 qubits active-space Hamiltonians. The effect of noise has the same behavior with $R_1$ than in 8 qubits case (Fig \ref{bench_noise_dist1} in Section \ref{sec:level27}), although the order of magnitude is larger. Not all values are available due to interoperability issues.}
     \label{bench_noise_dist1_16qb}
\end{figure*}

\clearpage

\nocite{*}

\newpage
\bibliography{bibli}

\end{document}